\documentclass[10pt,journal,compsoc]{IEEEtran}

\usepackage{tikz}
\usepackage{pgfplots}
\usetikzlibrary{arrows,arrows.meta,external,colorbrewer,decorations.markings,decorations.pathmorphing,decorations.pathreplacing,external,fit,patterns,shapes,positioning}
\usepgfplotslibrary{groupplots,colormaps,colorbrewer,statistics,fillbetween}
\usepackage[outline]{contour}
\usepackage{pgfplotstable}
\pgfplotsset{compat=1.15}
\usepackage{tcolorbox}
\usepackage[T1]{fontenc}
\usepackage{pgfpages}

\usepackage{tabularx}
\usepackage{enumitem}
\usepackage{adjustbox}
\usepackage{makecell}
\usepackage{subfig}
\usepackage{multirow}
\usepackage{textcomp}
\usepackage{bm}
\usepackage[absolute,showboxes]{textpos}

\usepackage{algorithmic}
\usepackage{graphicx}
\usepackage{xcolor}
\usepackage{hyperref}
\usepackage{booktabs}
\usepackage{tikz}
\usepackage{array}
\usepackage{balance}
\usepackage{microtype}
\ifCLASSOPTIONcompsoc
\usepackage[nocompress]{cite}
\else
\usepackage{cite}
\fi

\ifCLASSINFOpdf
\else
\fi

\newcommand{\mthree}{\textit{M3}\xspace}\newcommand{\esp}{\textit{ESP32}\xspace}\newcommand{\hifive}{\textit{HiFive}\xspace}

\usepackage{pifont}

\usepackage{xspace}
\newcommand{\cf}{\textit{cf.}~}
\newcommand{\etal}{\textit{et al.}}
\newcommand{\eg}{\textit{e.g.,}~}
\newcommand{\ie}{\textit{i.e.,}~}

\newcommand{\one}{({\em i})\xspace}
\newcommand{\two}{({\em ii})\xspace}
\newcommand{\three}{({\em iii})\xspace}
\newcommand{\four}{({\em iv})\xspace}
\newcommand{\five}{({\em v})\xspace}

\let\orgautoref\autoref
\renewcommand{\autoref}
{\def\sectionautorefname{Section}\def\subsectionautorefname{Section}\def\subsubsectionautorefname{Section}\orgautoref}

\makeatletter
\renewcommand{\paragraph}[1]{\vspace*{0.03in}\noindent{\bf #1.}\hspace{0.25ex \@plus1ex \@minus.2ex}}
\renewcommand{\subsubsection}[1]{\vspace*{0.03in}\noindent{\bf #1.}\hspace{0.25ex \@plus1ex \@minus.2ex}}
\newcommand{\paragraphS}[1]{\vspace*{0.03in}\noindent{\bf #1}\hspace{0.25ex \@plus1ex \@minus.2ex}}
\makeatother

\begin{document}
\title{PUF for the Commons: Enhancing Embedded\\ Security on the OS Level}

\setlength{\TPHorizModule}{.9\textwidth}
\setlength{\TPVertModule}{\paperheight}
\TPMargin{5pt}
\begin{textblock}{1}(.15,0.01)
  \noindent
  \footnotesize
  If you cite this paper, please use the IEEE reference:
  P. Kietzmann, T. C. Schmidt, and  M. W\"ahlisch.\\
  PUF for the Commons: Enhancing Embedded Security on the OS Level.
  \emph{IEEE Transactions on Dependable and Secure Computing}, IEEE Press : Piscataway, NJ, USA, 2023.
  \url{https://www.doi.org/10.1109/TDSC.2023.3300368}
\end{textblock}

\author{Peter Kietzmann,
        Thomas C. Schmidt,~\IEEEmembership{Member,~IEEE,}
        and~Matthias W\"ahlisch,~\IEEEmembership{Member,~IEEE}\IEEEcompsocitemizethanks{
\IEEEcompsocthanksitem Peter Kietzmann and Thomas C. Schmidt are with the Department Informatik, HAW Hamburg, Berliner Tor 7, 20099 Hamburg, Germany.\protect\\
E-mail: \{\texttt{peter.kietzmann, t.schmidt}\}\texttt{@haw-hamburg.de}.
\IEEEcompsocthanksitem Matthias W\"ahlisch is with the Faculty of Computer Science, TU Dresden, Helmholtzstr. 10, 01069 Dresden, and also with the Barkhausen Institut, 01187 Dresden,  Germany. \protect\\
E-mail: \texttt{m.waehlisch@tu-dresden.de}}.
% \thanks{Manuscript received Jan. 17, 2023.}
}

\markboth{}{}

\IEEEtitleabstractindextext{\begin{abstract}
Security is essential for the Internet of Things (IoT).
Cryptographic operations for authentication and encryption commonly rely on random input of high entropy and secure, tamper-resistant identities, which are difficult to obtain on constrained embedded devices.
In this paper, we design and analyze a generic integration of physically unclonable functions (PUFs) into the IoT operating system RIOT that supports about 250 platforms.  Our
approach leverages uninitialized SRAM to act as the digital fingerprint for
heterogeneous devices.
We ground our design on an extensive study of PUF performance in the wild, which involves SRAM measurements on more than 700 IoT nodes that aged naturally in the real-world. We quantify static SRAM bias, as well as the aging effects of devices and incorporate the results in our system. This work closes a previously identified gap of missing statistically significant sample sizes for testing the unpredictability of PUFs.
Our
experiments on COTS devices of 64\,kB SRAM indicate that secure random seeds derived from the SRAM PUF provide 256\,Bits-, and device unique keys provide more than 128\,Bits of security.
In a practical security assessment we show that SRAM PUFs resist moderate attack scenarios, which greatly improves the security of low-end IoT devices.
\end{abstract}

}

\maketitle
\section{Introduction}\label{sec:introduction}

\IEEEPARstart{T}{he Internet} of Things (IoT) comprises billions of constrained devices, but the low-cost IoT~hardware is challenged by basic security operations.
High entropy seeds for secure random number generation~\cite{ksw-gpngi-22} and secure hardware identities form the minimal set of primitives that bootstrap the cryptographic subsystem needed for protecting basic services of networked nodes.
These numbers must remain secret to prevent information leakage of past and future transactions, and require resistance against readout or tampering.
Supplementary hardware security modules (\eg secure elements) can overcome these challenges but increase device cost.
In practice, many large-scale IoT~deployments consist of cheap embedded~devices  without hardware security features, and readily threaten the IoT~\cite{kscga-atcai-19} as well as the global Internet~\cite{aabbb-umb-17}.

Physical unclonable functions (PUFs) utilize intrinsic hardware variations, which are a promising source of  \one random variations on one device, and \two unpredictable secrets between devices that become reproducible by excluding the variations from \one.
A prevalent type of PUF~input is SRAM. After powering up the hardware, SRAM provides a digital fingerprint from the patterns of uninitialized memory. SRAM is available on almost every IoT platform and can be exploited without additional hardware. This makes the technology particularly attractive for low-cost devices.
Secret values are generated only during system startup and consumed quickly after to lower the risk of a compromise. Consequently, SRAM secrets remain absent during regular node operations.

\begin{figure}[]
    \centering
    \includegraphics[width=0.675\columnwidth]{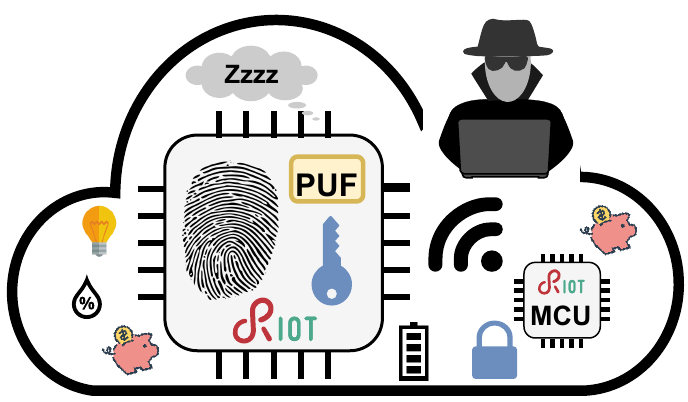}
    \caption{PUF security services provided by an operating system  enable lightweight crypto-operations on low-cost hardware in the IoT.}
    \label{fig:overview}
\end{figure}

There have been concerns, though, that the  physical layout of SRAM~\cite{mcms-lscr-10} as well as hardware aging~\cite{kks-insrs-06} may introduce systematic biases.
 Quantifying these subtle statistical effects requires a comparative analysis between large quantities of devices~\cite{wp-cbmpu-19}, which we contribute in this paper. Our large-scale evaluation of more than 700 nodes clearly shows a localized bias for certain bits in the SRAM response.
An attacker who tries to predict this pattern by using a large number of measurements from similar devices could reach an advantage in guessing the bit values at certain positions.
We quantify the remaining entropy of secrets derived by these biased pattern and identify secret generation schemes that are able to mitigate this weakness.

We further analyse hardware aging by extensive measurements on nodes that naturally aged in the real-world environment of an open access testbed. We find address specific wear-out effects that link to past experiment executions.
Our findings shall motivate testbed operators as well as software developers to invoke anti-aging strategies to their firmware.

In this paper, we design and evaluate PUFs for the multi-purpose operating system RIOT~\cite{bghkl-rosos-18} for constrained IoT devices~\cite{RFC-7228}.
Our configurable integration enables \one portability across heterogeneous platforms with differing hardware capabilities, and \two the replacement of modular building blocks in order to adapt to differing environmental conditions, security requirements, or the energy availability on battery driven nodes.
To the best of our knowledge, a consistent PUF integration into a commodity IoT operating system is yet missing, even though IoT deployments increasingly rely on some (open source) operating system (OS).
 IoT applications built on top of an OS benefit from reduced implementation overhead and enhanced dependability as they reutilize existing, well-tested code such as  network stacks, drivers, or crypto-libraries.

We argue that  operating system software should provide crypto-primitives for PUF functions  (see \autoref{fig:overview}) to make its security benefits accessible to a wide range of IoT devices.
The application of software PUFs is not limited to low-cost platforms, though, but can also assist security hardware, which is occasionally vulnerable~\cite{thn-crngf-21}.

The remainder of this paper is structured as follows.
After providing background on PUFs and discussing prior work~(\S\ref{sec:background}), 
we present our evaluation methodology (\S\ref{sec:setup_hw}) and perform an empirical SRAM evaluation on 708 devices (\S\ref{sec:eval_mem}), in which we identify static bias and stress marks introduced by past utilization.
We derive requirements for supporting SRAM PUFs on the OS-level, and introduce the RIOT OS integration  (\S\ref{sec:integration-riot}).
In a base-line evaluation of our solution  (\S\ref{sec:eval_cond}), we quantify the uniqueness of SRAM PUF generation.
Our second analysis concentrates on the quality of random seed and key generation, as well as its performance overhead (\S\ref{sec:eval_seed_key}).
A subsequent security analysis reveals that the SRAM PUF is secure under moderate attacker assumptions (\S\ref{sec:secu_consider}).
Finally, we conclude and outline future research directions (\S\ref{sec:conclusion-outlook}).

 \section{Problem Statement and Related Work}\label{sec:background}

Pappu~\etal~\cite{prtg-pof-02} are the first to introduce ``physical one-way functions'' and the notion of a PUF dates back to
Gassend~\etal~\cite{gcdd-sprf-02}; both describe a technique to uniquely identify and authenticate individual integrated circuits.
The research community identified PUFs as an attractive solution for the IoT~\cite{snffi-csitd-18}, because the intrinsic hardware variations can feed security  primitives on low-end devices without increasing hardware cost.
Orthogonal to PUFs that utilize variations of low-cost, multi-purpose building blocks, research also advances in the field of hardware security at the transistor level~\cite{smsak-dpufc-17,tra-iutmp-22,hlyy-azber-23}.
PUFs can be distinguished into two classes~\cite{prtg-pof-02, gkst-fiptu-07}. They either process many inputs (\ie challenges) to produce varying outputs (\ie responses), or only few inputs which produce few, or only one response.
More precisely, a PUF is denoted as strong if it provides a large challenge-response space, whereas a weak PUF provides only few challenge-response pairs that typically scale linearly or polynomially with the design size~\cite{mbwry-pt-19}.
Hence, a memory readout which produces one response can be classified as a weak PUF.

A variety of use cases for PUFs emerged, such as
secure key storages~\cite{ekl-lrpms-11},
communication protocols~\cite{ccm-pscpi-17,brr-rcaii-21},
supply chain security~\cite{ffgrs-pstsc-21},
remote attestation~\cite{ssw-splra-11},
firmware updates~\cite{pvb-tfuid-17}, or generic trust anchors~\cite{h-ata-21}.

The security of these applications as derived from PUFs is only as strong, as the secrets extracted from the underlying hardware variations. In this section, we review the fundamental properties, basic assessment measures, and pose the question of potential weaknesses in PUFs.

\subsection{Properties of Uninitialized SRAM}\label{sec:back_mem}
Reading out uninitialized SRAM produces a digital fingerprint.
Manufacturing processes introduce variations in the silicon of transistors that construct a memory cell.
When powered on, some cells drift to the logical state 1, others to 0, and  cells without bias fluctuate according to environmental conditions.
The resulting patterns require a careful assessment between devices (inter-device) in order to estimate their uniqueness, and between power-cycles on one device (intra-device), in order to quantify the  (random) noise. This noise can be utilized as an entropy source, or needs to be removed for reliably reproducing an exact version of the pattern.

\paragraph{Aging Bias}
Uniformly random variations result in an equal proportion of stable cells that power up with 0 or 1 on a single SRAM pattern.
Aging and utilization, however, skew this distribution, due to drifting voltage threshold values of the transistors that form a memory cell~\cite{fn-atopd-14}.
The increased probability for one symbol (1 or 0) introduces a bias, which in turn benefits an attacker, who tries to guess bit values.
Guin~\etal~\cite{gwhs-drsea-19} find a bias of up to 54\,\% under artificial aging. 
Holcomb~\etal~\cite{hbf-pssif-09} counter that `normal' use patterns of intermittently powered devices prevent an identical skew.

The state-of-the-art motivates additional and more realistic analyses of bias and aging effects on platforms that naturally aged while executing real-world IoT applications over a long period of time (see \autoref{sec:eval_mem}).

\paragraph{Static Bias}
In contrast to an aging bias,
real-world PUFs may be affected by a static bias at certain bit positions~\cite{wfp-ebcmp-19}.
Rahman~\etal~\cite{rhgcf-sccna-17} observe systematic correlation between SRAM patterns across chips, and cell-neighborhood interactions due a systematic physical arrangement on the silicon.
Both effects reduce the device uniqueness.
An attacker, who owns a chip of the same type, could utilize a local measurement to guess bit values at specific positions with a better chance than 50\,\%, which facilitates prediction of a secret value derived thereof.

Conditioning the SRAM resolves systematic bias.
Bit selection~\cite{xrfhs-bsash-14, kacp-pcspu-19} is an approach to exclude biased bit addresses of the SRAM, but adds enrollment complexity for each individual node. Storing a bit mask for cell selection requires additional memory, which conflicts with limited memory resources on IoT devices.
Instead, extending the SRAM PUF input increases the total amount of unbiased bits that generate a secret, which ideally prevents successful guesswork.
Increasing the length, though, threatens \textit{fuzzy extraction} (see~\autoref{sec:back_secu}) which may leak information about the secret, and requires a careful assessment of \one the remaining entropy as well as \two the increase in processing overhead on resource constrained nodes.

\paragraph{Environmental Bias}
PUFs are subject to noise, which is affected by environmental operating conditions.
Related work analyzes SRAM startup patterns under varying voltages and temperatures.
A body of work shows that SRAM PUFs are robust against variations of the supply voltage~\cite{sl-casmu-12,kkrsv-pmfbs-12,clb-csfpt-12,bbmm-tmspq-15}, tested at $\pm$\,10\,\% of the nominal value.
Adjusting the voltage ramp-up speed~\cite{chlms-avrtt-13, ljj-pctrs-19} can mitigate effects of temperature variations, which affect the SRAM startup behavior more severely. These solutions require special hardware, though.

The effect of temperature variations on the startup pattern of SRAM is commonly analyzed within the industrial operating temperature range of -40\textdegree C to +80\textdegree C.
Leest~\etal~\cite{lssth-eitrn-12} quantify the \textbf{intra-device} relation between startup pattern of a device, which shows that the min. entropy is minimal at a low temperature of -40\textdegree C, and gains up to $\approx$2\,\% of noisy cells when the temperature increases to +80\textdegree C.
When quantifying the required SRAM length for seed generation, the lower bound should serve as a conservative starting point, while higher temperatures improve seed generation due to higher entropy.

Schrijen~\etal~\cite{sl-casmu-12} compare the intra-device hamming distance of SRAM patterns taken under different temperatures, compared to an enrollment readout at ambient temperature.
The startup noise between patterns of the same device almost doubles when increasing the operation temperature from 20\textcelsius ~to 80\textcelsius.
Claes~\etal~\cite{clb-csfpt-12} present an increase of the fractional hamming distance from 0.06 to 0.1 at the extreme operating temperatures of -40/+80\textdegree C.
Katzenbeisser ~\etal~\cite{kkrsv-pmfbs-12} present similar results but find that the SRAM PUF is more robust against varying operational conditions compared to other PUFs (\eg arbiter or flip-flop PUFs). 
Overdesigning the error correction code can mitigate
this effect, but increases the computational complexity— sometimes in
conflict with IoT device constraints as well as the remaining key entropy.

Holcomb~\etal~\cite{hbf-pssif-09} show that the increase in noise which is introduced by temperature variations overrules a predictable aging effect of NBTI.
This has a positive effect on the \textbf{inter-device} uniqueness, which increases with the absence of an identical skew.

\subsection{Empirical Evaluation of PUFs}\label{sec:back_large}
The common measure to quantify the unpredictability of a pattern is given by the min. entropy metric:

\begin{equation}
H_{min}(p_{max}) = -log_2(p_{max})
\label{eq:hmin}
\end{equation}

\noindent For a single bit, $p_{max}=max(p, 1-p)$, \ie the maximum probability for attaining one ($p$) or zero ($1-p$) at the same SRAM bit position. An ideal probability of $p_{max}=0.5$ maximizes the min. entropy to $H_{min}=1$.
This metric is used to assess intra-device variations across multiple pattern of the same device, or inter-device variations between the pattern of multiple devices.
Random noise increases the intra-device min. entropy after a power-cycle, which facilitates seed generation, but challenges a reliable key construction. Over-dimensioning the \textit{fuzzy extractor} (see \autoref{sec:back_seed_key}) can mitigate this effect, but increases the computational complexity---in conflict with IoT device constraints.

Schrijen~\etal~\cite{sl-casmu-12} present intra-device measurements across SRAM technologies in differing setups and find that variations across SRAM of different vendors are not significant. 
Katzenbeisser~\etal~\cite{kkrsv-pmfbs-12} show that the inter-device min. entropy is invariant to temperature. This enables longer repetition codes to correct multiple errors, which would otherwise leak secret information in the case of a low inter-device entropy (\cf~\autoref{sec:eval_seed_key}).

The inter-device min. entropy assesses device uniqueness and the impact of bias.
The literature reports inter-device min. entropy values from 0.7~\cite{clb-csfpt-12} to 0.9~\cite{klrw-elpkg-14} between SRAM patterns. Quantifying this metric requires multiple samples which is particularly challenging since it involves many nodes.

\paragraph{Min. Entropy Convergence}
The maximum probability $p_{max}$ in \autoref{eq:hmin} can be empirically sampled from a limited number of probes $n$ (\ie  nodes). Then the empirical estimator 

\begin{equation}
H^\prime_{min}(i, n) = -log_2\left[max\left(\frac{i}{n}, 1- \frac{i}{n} \right)\right]
\label{eq:hmin_i_n}
\end{equation}

\noindent with  $i$ positive events (ones)  in $n$ samples from individual nodes 
converges to the min. entropy in \autoref{eq:hmin}. Statistical convergence, however, is slow. According to the central limit theorem~\cite{f-iptia2-71},

\begin{equation} 
	\left| H_{min}(p_{max}) - H^\prime_{min}(i, n) \right| \sim \frac{\sigma}{\sqrt{n}}\ \ (\textrm{as}\, \,n \rightarrow \infty),
    \label{eq:se_hmin}
\end{equation}

\noindent where the dispersion $\sigma = \sigma_{H^\prime_{min}} \approx 1$.

Hence, estimating the inter-device bias from 100 samples of SRAM PUFs still includes an error of 10\%. Accordingly, the largest available SRAM evaluation of 144 nodes~\cite{w-lscsi-17} bears an uncertainty of more than 8\%   
This shows the need for significantly larger samples in order to approximate the inter-device min. entropy accurately, which we will present in \autoref{sec:eval_mem}.

\paragraph{Bit-Aliasing}
Maiti~\etal~\cite{mcms-lscr-10} introduce \textit{bit-aliasing} to quantify systematic inter-device bias (cf. \autoref{sec:back_mem}) among 125 FPGAs that implement a ring oscillator (RO) PUF.
Large-scale evaluations of RO PUFs on 217 FPGAs~\cite{hwgh-lsrpa-18,gclhm-lcesr-20}, and 133 ASICs~\cite{yssmd-pmerp-12} show that the location of cells within the FPGA affect performance properties.
The bit-alias of uninitialized SRAM between 50~\cite{bbmm-tmspq-15} and 144~\cite{w-lscsi-17} devices reveals a slight double-peaked distribution of the bit-alias scores due to SRAM layout systematics, but seem to miss convergence due to an  insufficient sample size.Wilde~\etal~\cite{wp-cbmpu-19} identify a research gap in convergence and deduce that qualified inter-device bit-alias measurements require more than 600 devices to converge with an error below 5\%.

Quantifying possible inter-device correlations using  hundreds of devices demands for high cost and engineering efforts. In the subsequent analyses, we will tackle these challenges by taking advantage of a large-scale testbed.

\subsection{Random Seed and Key Generation}\label{sec:back_seed_key}
SRAM PUFs promise to support bootstrapping security on embedded IoT nodes by deriving random seeds and private keys from uninitialized memory.
Commercial IoT platforms more and more provide isolated PUF circuits for this purpose, but an open software implementation that enables PUF-functionality without dedicated PUF-circuitry is missing.
To enable software-based SRAM PUFs on a wide range of heterogeneous IoT platforms, the hardware abstraction layer of an operating system can enable low-level hardware access and facilitate PUF-based seed and key generation.

\paragraph{Seed Generation}
Random numbers are essential for security. Commonly, a sequence generated by a true random number generator (TRNG) acts as seed or refresh value for a pseudo-random number generator (PRNG) as well as a cryptographically secure PRNG (CSPRNG). Van der Leest~\etal~\cite{lssth-eitrn-12} derive the min. entropy of repeated SRAM startup patterns on a device for creating a random seed value. The concept was applied to off-the-shelf MCUs~\cite{hlskv-spsco-13} and revealed a diverse picture. Not all embedded SRAM technologies are qualified to produce high entropy seeds. SRAM, so the lessons learned from this study, must be analyzed prior to deployment.
Krentz~\etal~\cite{kmg-sswps-17} propose an SRAM seeding mechanism and add antenna noise to uninitialized memory pattern. The combined values are conditioned with a van Neumann extractor, which introduces variable runtime overhead.

SRAM must be uninitialized to obtain entropy between power-cycles, which is why the PUF operation should only take place during system startup before the memory has been utilized. This startup sequence, however, might be executed without a cold boot, possibly leading to zero-entropy seeding. Hence, a PUF implementation needs to ensure a preceding power-off cycle.

\paragraph{Key Generation}
A reliable key generation depends on the removal of random noise. The related concept of \textit{fuzzy extraction} was first presented in the context of biometric authentication systems~\cite{jw-fcs-99,dors-fehgs-08} to reliably reconstruct an exact version of a reference measurement.
Fuzzy extractors are based on error correction codes.
Error correction schemes for PUFs~\cite{yd-srecp-10} were evaluated on an FPGA~\cite{hks-recpe-20}. For complexity reasons, not all codes are applicable to low-end devices with its constrained resources. Korenda~\etal~\cite{kacp-pcspu-19} reduce the computational requirements by identifying stable values before encoding, which reduces the error probability.
Leest~\etal~\cite{lps-sdecc-12} propose specific hardware implementations for soft-decision decoders, which improve the correction capabilities and require only half of the PUF bits for secret generation compared to hard decision decoders.

A deployment of a fuzzy extractor proceeds in two phases,  \textit{enrollment} and \textit{reconstruction}.
The enrollment is a trusted process and produces \textit{helper data}~\cite{dgsv-hdapk-15}, which is later used to reconstruct the PUF value. Helper data is publicly stored in non-volatile memory. 

A PUF response does not contain maximum entropy, which flaws its immediate use as a cryptographic key.
Besides, it may be too long or too short, adjusting its length to include a required amount of entropy.
For mitigation, a compression scheme can be used to create a key with maximized entropy and to preserve forward secrecy of the PUF response. Practical implementations~\cite{ed-pufda-07, bgsst-ehdke-08, mhv-pffpc-12} employ a cryptographic hash function that compresses the lengthy PUF response.

Error correction~\cite{dgsv-hdapk-15}, and crypto-processing~\cite{kblsw-pscli-21} quickly exceed the computational-, and energy resources on constrained embedded devices.
A modular and configurable PUF implementation should ease the deployment under varying environmental conditions and adjust to the capabilities of heterogeneous platforms (\eg processing power, availability of crypto-acceleration).

\subsection{Security Analysis of PUFs}\label{sec:back_secu}

The related work presents threats to PUFs mainly from three angles.

\paragraph{Analytical Attacks}
Public helper data techniques leak information if the PUF is biased~\cite{mmdv-srpsc-13}. For the \textit{code offset} method, helper data lengths should be kept small to avoid information disclosure--in conflict with PUF bias which may increase the required length. Koeberl~\etal~\cite{klrw-elpkg-14} conservatively estimate the entropy loss during helper data construction for varying error correction codes in the fuzzy extractor, but were criticized to be overly pessimistic~\cite{dgsv-hdapk-15}. Maes~\etal~\cite{mlsw-skgbf-16} present methods that calculate the entropy leakage exactly, and de-biasing which resolves bias on an FPGA.
Liu~\etal~\cite{llltl-mecim-18} present countermeasures to bias on an MCU.

\paragraph{Modeling Attacks}
PUFs are susceptible to modeling attacks~\cite{gtfs-smlaa-16,slz-aasp-20,wtmsz-nnmaa-22}. R\"{u}hrmair~\etal~\cite{rssdd-mapuf-10} apply machine learning to challenge-response pairs of PUFs with many inputs and predict their outputs, which requires the ability to eavesdrop PUF responses.
Strieder~\etal~\cite{sfp-mlpuf-21} exploit helper data of PUFs with many inputs for training.
PUFs with few (or only one) input are less vulnerable to learning attacks due to restricted input/output variables.

\paragraph{Hardware (Invasive) Attacks}
Helfmeier~\etal~\cite{hbnf-cpuf-13} cloned SRAM of a common IoT device using a focused ion beam instrument.
Zeitouni~\etal~\cite{zowks-rdspc-16} present a side-channel analysis on an SRAM PUF, using remanence decay. Both attacks require physical control of the device under attack.

These analyses are tied to specific algorithms, or dedicated PUF implementations in hardware or software.
A practical threat model that analyses the remaining security risks of SRAM PUFs on low-end hardware from the perspective of an IoT operating system is missing. We will fill this gap in \autoref{sec:secu_consider}.

 \section{Experiment Setup}\label{sec:setup_hw}

We want to analyze the properties of uninitialized SRAM on a large scale, and assess our measurements on IoT-typical constrained hardware.
Therefore, we chose an existing testbed as an evaluation environment (\autoref{sec:meth_testbed}), which provides many off-the-shelf nodes (\autoref{sec:meth_hw}) and grants open (remote) access for reproducibility.
The drawback of this approach, however, is that we cannot vary the operational conditions of nodes.

On the software side of our experiments (\autoref{sec:meth_sw}), we chose the open source IoT operating system RIOT~\cite{bghkl-rosos-18} for three reasons.
\one RIOT is an off-the-shelf OS that is used in many IoT deployments~\cite{eclipse-iedsr-19} with support of numerous heterogeneous platforms. In RIOT, PUF support brings benefit to a broad range of systems and applications.
\two It provides support for the FIT IoT-LAB testbed nodes. This allows us to easily benefit from the existing tools and facilities. 
\three An active open source community, which had first hand experiences with initial SRAM PUF trials~\cite{kgsw-psgri-18}, facilitates code contributions.

\subsection{Testbed Environment}\label{sec:meth_testbed}
We conduct our experiments on the FIT IoT-LAB testbed~\cite{abfhm-filso-15} to attain a large number of nodes. The testbed consists of seven sites with different topologies and a total number of more than 1500 nodes of 25 architectures. The \mthree nodes make up the majority ($\approx$\,800 nodes) and reflect properties of commercial off-the-shelf class~2 IoT devices~\cite{RFC-7228}.
Each node is attached to a control node which provides a power monitor (INA220), allowing to measure the operational voltage and current that flows to the MCU and the external board components.
Nodes are deployed across facilities of INRIA in France. Hence, all our experiments are conducted under environmental conditions of work offices.

We use 708 \mthree nodes in our experiments.
To automate experiment control, we utilize the command-line interface \texttt{iotlabcli}. Nodes serial outputs are piped to individual log files.
Note, when reproducing the experiments, high data volumes are generated, while testbed users have limited disk quota. Data compression, moving files periodically, and asking for increased quota can assist.

\subsection{Hardware Platform}\label{sec:meth_hw}
\paragraph{Testbed}
\mthree nodes consist of a 32-bit ARM Cortex-M3 CPU, integrated into the STM32F103REY MCU, which runs at max. 72\,MHz and provides 64\,kB embedded SRAM, and 512\,kB internal flash.
The MCU offers common features that we exploit:
\one Low-power standby mode turns off the whole SRAM. All content in SRAM and registers are lost, except for the backup domain.
\two Real-time clock remains operable during standby, to trigger an interrupt for wakeup.
\three Power control registers indicate whether the MCU has been in standby after a system restart.
But this MCU lacks hardware security features, \ie a random number generator, crypto-accelerator, and secure key storage.
\mthree nodes additionally connect external components via SPI: An 16\,MB external NOR flash allows storing data persistently, and a low-power radio which is the only alternative to gather entropy on this board, by sampling antenna noise.

The microchips of the \mthree nodes in the FIT IoT-LAB testbed originate from two lots and four wafers, two of which build the majority of devices. We conducted several experiments to find a systematic variation. But we could not find significant differences between these batches, hence, we treat them equally in our evaluation and exclude the results of the batch comparisons.

\paragraph{Local} To evaluate the PUF performance on heterogeneous IoT devices with varying architectures, we also perform local experiments on two different off-the-shelf IoT platforms, and measure the processing time on a single device per platform:
The \esp, which consists of an Xtensa 32-bit CPU with 520\,kB SRAM, 4\,MB flash, and operates at max. 240\,MHz.
The \hifive which consists of a RISC-V RV32IMAC CPU that provides 16\,kB SRAM, 4\,MB off-chip flash, and operates at max. 320\,MHz.

\subsection{Software Platform}\label{sec:meth_sw}
We base our PUF implementation on RIOT 2022.01. It supports different architectures (8--32-bit CPUs), over 150 MCUs, and nearly 250 IoT boards.
The OS provides multi-threading with preemption, power management, and a
hardware abstraction layer to enable portability. We utilize and complement these features in our implementation (\autoref{sec:integration-riot}).
RIOT provides its own IPv6 network stack (\texttt{GNRC}) and supports multiple low-power radios as well as wired interfaces.
For the \mthree nodes,  we added drivers to access power control registers and the external flash memory. To broaden our experimental basis, we integrated the PUF initialization to the  \esp and \hifive architectures.

In our experiments, we trigger repeated power cycles on the nodes.
For this, we utilize  existing RIOT interfaces, namely, the power management (PM) interface to enter standby, which turns-off the SRAM, and the real-time clock (RTC) to generate a future wakeup interrupt.
 \section{Large Field Study of Uninitialized SRAM}\label{sec:eval_mem}

\begin{figure}[]
    \centering
    \includegraphics[width=.85\columnwidth]{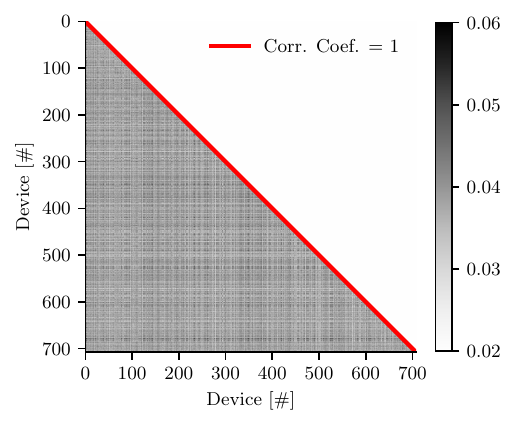}
    \caption{SRAM correlation between 708 nodes. The Pearson product-moment correlation coefficient of each pair is encoded in gray intensity. Autocorrelation results in a coefficient of one.}
    \label{fig:sram_corrcoef}
\end{figure}

\subsection{Inter-device Correlation}\label{sec:eval_mem_similar}
We want to analyze the similarity between individual SRAM patterns. Therefor we read the whole memory of 708 available \mthree nodes and compute the Pearson product-moment correlation coefficient which is defined as:

\begin{equation}
    r(a, b) = \frac{\sum_{i=1}^{m} (a_i - \bar{a})(b_i-\bar{b})}{\sqrt{\sum_{i=1}^{m} (a_i - \bar{a})^2 \vphantom{(\bar{b})^2}} \sqrt{\sum_{i=1}^{m} (b_i - \bar{b})^2} }
    \label{eq:correlation}
\end{equation}

\noindent  where $a$ and $b$ denote the SRAM pattern of two devices with a length of $m$=64\,kB.
\autoref{fig:sram_corrcoef} presents the matrix of correlation coefficients between the SRAM readout of all node pairs, as a measure of linear dependency between nodes.
A coefficient of 1 indicates perfect correlation (pairs are equal), -1 represents negative correlation (pairs are opposite), and 0 means (linear) independence.
All coefficients are small with a small positive bias (0.02--0.06), which indicates high independence between the memory patterns and motivates their usage as PUF source.  Certain samples, however, indicate a slightly increased coefficient when compared to others.
To better understand these correlations, we chose to further analyze the inter-device relations with a metric that incorporates the bit locality, \eg the bit-alias~\cite{mcms-lscr-10} quantifies inter-device bias (\cf~\autoref{sec:eval_alias}).

\subsection{Analysis of Static Bias}\label{sec:eval_alias}
\begin{figure}
    \centering
    \includegraphics[width=1\columnwidth]{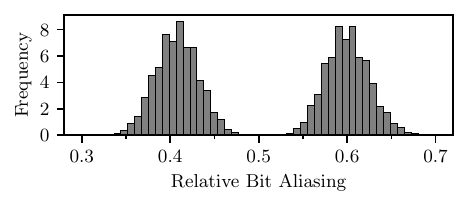}
    \caption{Distribution of bit-alias values between 708 nodes.}
    \label{fig:hist_wo_shuffle}
\end{figure}

We calculate the bit-alias which quantifies the proportion of zeros and ones at every bit position $j$ in the memory pattern between $n$ devices:

\begin{equation}
    \hat{p}_{j} = \frac{1}{n} \sum_{i=1}^{n} {p}_{j, i},
    \label{eq:bitalias}
\end{equation}
where ${p}_{j, i}$ denotes the measured bit probability at position $j$ of device $i$. If the probability for attaining one or zero is unbiased, the expectation value at each bit position equals $0.5$ and the distribution of the empirical $\hat{p}_j$-values follows a normal (error) distribution.
Our evaluation indicates repetitive pattern with (multiples) of 32\,Bit blocks.
Rahman~\etal~\cite{rhgcf-sccna-17} find similar effects and relate this to the physical layout of the SRAM.
\autoref{fig:hist_wo_shuffle} displays the histogram of the bit-alias metric for all bit positions of the 64\,kB memory. It reveals a bimodal distribution with peaks around 0.4 and 0.6.
Previous work~\cite{w-lscsi-17} suggested a double-peak distribution, but its sample size was too small~\cite{wp-cbmpu-19}. To the best of our knowledge, our results show the first  SRAM evaluation of the bit-alias with an error around 3\,\%.

Inter-device correlations in regions of SRAM can be beneficial for an attacker. Analysing a large set of equally produced devices may assist prediction of SRAM bit values at certain positions.
In detail, a deviation of $\approx$\,0.1 from the ideal bit probability of $p_j=0.5$ increases the chance of guessing the correct value  by 10\,\%.
This lowers the inter-device entropy, which we quantify in~\autoref{sec:eval_entropy}, and requires careful consideration when generating keys (see~\autoref{sec:eval_keygen}).
While not all SRAM technologies seem to be affected by this inter-device correlation, a pre-selection of uncorrelated bits for the enrollment process can mitigate this effect~\cite{rhgcf-sccna-17}.

\subsection{Analysis of Aging}\label{sec:eval_mem_bias}

\begin{figure}[]
    \centering
    \includegraphics[width=1\columnwidth]{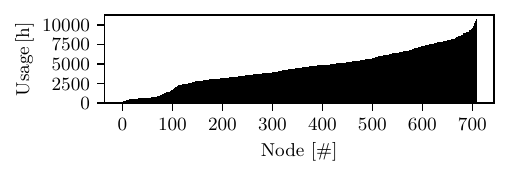}
    \caption{\mthree node active experiment operation time in hours. Nodes are ranked according to their utilization.}
    \label{fig:utilization}
\end{figure}

The MCU age is noted on the chip package and our local \mthree sample devices indicate a production date in January 2012.
This is in line with testbed statistics that date back the first experiment to September 2012.
We further managed to get experiment metadata from the testbed team. \autoref{fig:utilization} displays the active utilization time of our test nodes since their deployment until the end of 2021.
The majority of our publicly accessible nodes have been operated 2.5--8 thousand hours since their deployment.
Thus, in contrast to prior work, we analyze devices that naturally aged under real-world conditions.

\begin{figure}[t!]
    \centering
    \begin{adjustbox}{max width=1\columnwidth}
        \subfloat[Avg. and 25-75th\,perc. weight by device.]{\includegraphics[height=.2\paperheight]{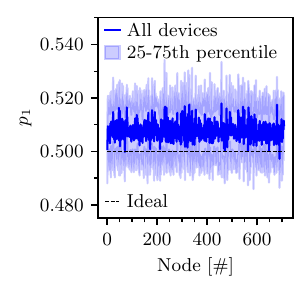}\label{fig:weights_compare_nodes}}
        \subfloat[Avg. weight by address.]{\includegraphics[height=.2\paperheight]{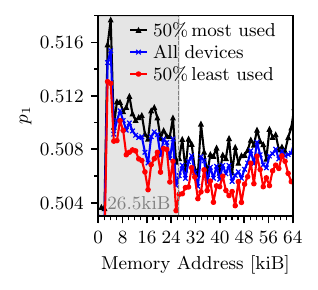}\label{fig:weights_compare_addresses}}
    \end{adjustbox}
    \caption{64 kB SRAM is split and analyzed in blocks of 1024 Bytes. The relative hamming
    weight is displayed for every device (\ref{fig:weights_compare_nodes}) and memory address (\ref{fig:weights_compare_addresses}); the latter distinguishes the half most/least used devices.}
    \label{fig:hamm_weight}
\end{figure}

We want to analyze whether certain devices or memory blocks show anomalous behavior caused by aging or wear-out from similar firmware images.
To that end, we quantify the intra-device bias by calculating the relative hamming weight:

\begin{equation}
    HW(r) = \left| \{r_i \neq 0 : 1 \le i \le m\} \right| \cdot \frac{1}{m}
    \label{eq:hamming}
\end{equation}

\noindent  where $r$ denotes the bit value of one device at position $i$ in a block of the length $m$. Hence, the hamming weight reflects the proportion of ones ($p_1$).
The proportion of ones and zeros should be equal ($p_1 = 1 - p_0 =0.5$) without bias.
Figure~\ref{fig:weights_compare_nodes} displays the intra-device measurements across the whole memory of all boards ($m$=64\,kB) which shows an average hamming weight of $0.508\pm 0.003 (\sigma)$. This slight (positive) bias is the effect of aging and is still small compared to the results of Guin~\etal~\cite{gwhs-drsea-19} who find biases of up to 0.54 after 336 hours (14 days) of stressed operation.

Figure~\autoref{fig:weights_compare_addresses} displays the hamming weight separated into memory blocks ($m$=1024\,Bytes). We show average values across all devices, and two subsamples that include 50\,\% of the most and least used devices.
\one An increase at $\approx$\,4\,kB is introduced by the bootloader. In real-world implementations, this can barely be avoided and PUFs should exclude SRAM at that region.
\two The bias of heavier utilized devices increases less used ones by $\approx$\,0.0025, which confirms aging by operation, with a small magnitude.
\three Besides \two, the first $\approx$\,26.5\,kB of memory exhibit a higher skew compared to the remaining.
Common firmware sizes of large-scale networking experiments on these testbed nodes report (\eg \cite{gklp-ncmcm-18}) memory requirements of 22--28\,kB in RAM, which matches the region of systematic wear-out. Hence, we report strong indications of visible wear-out effect by long-term testbed utilization and avoid that memory region in our PUF design (Sections~\ref{sec:integ_seeder} and~\ref{sec:integ_keygen}).
Operating systems likely organize the program memory from the start of the address space. At the same time, real- world firmware images do not necessarily utilize the whole memory (uniformly), which fosters bespoke unbalanced wear-out effects.
Testbed operators as well as PUF developers should include anti-aging techniques in the future  to mitigate wear-out patterns of various characteristics in practice.

 \section{PUF Design for the RIOT OS}\label{sec:integration-riot}
\begin{figure*}
\centering
    \includegraphics[width=1.0\textwidth]{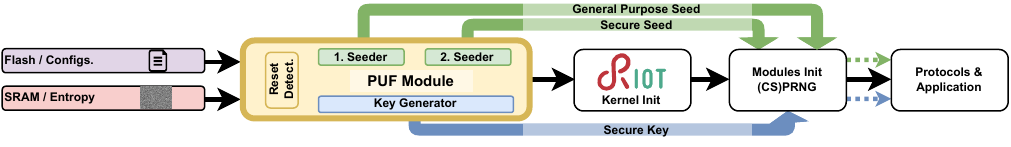}
    \caption{Integration of the SRAM PUF module in the IoT operating system RIOT.}
    \label{fig:puf_riot}
\end{figure*}

A wide availability of PUFs requires grounding in the ecosystem of an OS. The heterogeneity of supported platforms requires an integration into the configuration and the build system to adjust the diverse device properties. OS tests and tools provide useful interfaces to verify PUF viability and to assess crucial configuration parameters (\eg required SRAM lengths). PUFs bootstrap system security and must therefore extend the OS startup code, module initialization, and finally the secure operation.
\autoref{fig:puf_riot} presents an overview of our PUF integration in RIOT for creating
\one a simple seed for general purpose PRNG initialization,
\two a secure seed for CSPRNG initialization, and
\three a secret key.
In addition, to ensure qualified PUFs, we provide
a soft-reset detection mechanism that prevents initialized SRAM (\ie caused by insufficient power-off cycles) from generating seeds or keys.

\subsection{Compile-time Configuration}\label{sec:integ_config}
RIOT supports many boards of largely varying hardware capabilities~\cite{bklsw-usioi-22} that demand for a systematic compile-time modeling of its features. This modeling enables extensible code paths where possible, and facilitates reduced feature sets on platforms without certain hardware capabilities.
RIOT uses a feature modeling based on Kconfig~\cite{linux-kconfig-20, kblsw-pscli-21}.
Kconfig allows defining symbols that represent features, based on which dependencies and conditional default values are defined.
For the PUF module, a platform can indicate \textit{capabilities} as follows.
\texttt{HAS\_PM} enables low-power mode, and
\texttt{HAS\_PM\_TIMER} enables programmatic wake-up from low-power mode.
\texttt{HAS\_PM\_INDICATION} enables additional power-cycle detection during soft-reset detection.
\texttt{HAS\_CRYPTO\_ACCEL} enables crypto hardware acceleration (future work).

Both seeders (\autoref{sec:integ_seeder}) and the key generator (\autoref{sec:integ_keygen}) provide \textit{configurations} for PUF algorithms:
\one separate start addresses in SRAM,
\two length of the considered SRAM blocks,
\three choice of a cryptographic hash function,\four configuration of the error correction code for the key generator.
Default values are chosen according to our evaluation.

\subsection{Integration into OS Startup Routine}\label{sec:integ_startup}
\paragraph{System Reset}
RIOT provides a \texttt{reset\_handler}, which is the start point after every system reset. A default startup routine follows four steps. \one The data section is loaded from flash to RAM. \two~The \textit{.bss} (block starting symbol) section (used for uninitialized data) is set to zero. \three The MCU and board specific components are initialized. \four The OS kernel is loaded and \five \texttt{auto\_init} initializes modules prior to starting applications.We perform our PUF initialization prior to step \one, to obtain a pristine response of uninitialized memory.

\paragraph{Linker Attributes and Erasure}
To prevent PUF outputs from erasure by the subsequent startup routine, a \textit{.noinit} section in the linker script of every supported CPU architecture defines a PUF attribute with which we declare variables used to store the PUF seeds and keys.
Seeds and the key are consumed during \texttt{auto\_init} and unavailable by the end of the initialization (see \autoref{sec:ineg_access}).

\paragraph{Startup Delay}
PUF execution adds a delay (see~\autoref{sec:eval_time}) to the system startup, which is primarily introduced by resource intensive crypto-operations on constrained devices~\cite{kblsw-pscli-21}.
If available, crypto-accelerators can reduce that time.
When operated in software, the execution may degrade due to the early PUF execution on perhaps uninitialized system clocks, prior to MCU initialization.
An interface that allows conditional PUF execution during the next reset can mitigate this affect in the future.

\subsection{Detection of Soft Resets}\label{sec:integ_softreset}

Memory must be uninitialized for PUF operations, which is achieved by a sufficiently long power-off cycle. Short resets can occur, however. In such cases, the PUF procedure must not be executed to prevent duplicate seeds and false key construction. Kietzmann~\etal~\cite{ksw-gpngi-22} present a simple detection mechanism to catch soft-resets. In a nutshell, the soft-reset writes a memory marker to a known address.
On soft-reset, the marker will persist in memory.
Conversely, a sufficient power-off cycle changes the value of the marker and enables PUF operation.
One caveat of this approach is a false negative decision, which can be triggered by only a single bit flip in the marker variable, while old values stay in memory.
In a baseline experiment we analyzed the hamming distance between the marker variable and the expected value and decreased the duration of the low-power cycle.
Our results show notable signs of memory retention (decrease of the hamming distance) below 3\,ms, which should be excluded. Longer cycles, however, may still incur memory retention, sometimes in the order of seconds~\cite{sl-casmu-12}.

\paragraph{Distance Detection}
A future soft-reset detection stage could improve the false negative sensitivity and incorporate a distance metric to the detection algorithm.
Additionally, an individual marker per device could utilize the inverse startup pattern, maximizing the range of potentially flipping bits, which increases granularity.

\paragraph{Sleep State Report Interface}
The memory marker is at risk to be manipulated during runtime, by software defects, or intentionally by adversaries that manage to execute malicious code (\autoref{sec:secu_consider}).
This can cause an undetected soft-reset, resulting in zero-entropy seeding and false key reconstruction.
We extend the power management (PM) API in RIOT by a function to report the preceding state after a reset. Only if the marker-based detection fails \textit{and} the system starts from deep sleep, PUF operation is executed. Not all platforms support this feature, unfortunately.

\subsection{Random Seed Generation}\label{sec:integ_seeder}
Uninitialized SRAM contains randomness for the seed generator and is compressed to provide a concise value of maximized entropy.
We provision two seed generation functions that take as input the SRAM start address and considered memory length.
By default, we utilize a randomly chosen start address in the center of the memory map, to circumvent systematic wear-out effects that likely occur in the beginning of the RAM (\cf~\autoref{sec:eval_mem_bias}), and we locate regions for both seed functions successively.
Addresses and lengths can be configured, though. A dynamic mechanism could thus mitigate potential aging phenomena.

\paragraph{Construction}
Our first seed is extracted by the lightweight DEK hash~\cite{k-acp-09} and compressed to an integer value, which is utilized to seed a non-secure general purpose PRNG.
The second seed is created for security purposes and bases on compression by a cryptographic hash (SHA256 by default). Hence, the size of the seed corresponds to the digest length. It can be utilized to feed an entropy accumulator, or the CSPRNG initialization directly. Potential CSPRNG re-seeding~\cite{ksw-gpngi-22}, however, requires a power-cycle to obtain fresh entropy from the SRAM.

\paragraph{General Purpose vs Secure Seeds}
General purpose seeds must not be used in cryptographic contexts due to insufficient entropy and lack of forward secrecy. Conversely, cryptographic seeds can be used for general purpose, but exhibit higher cost (see~\autoref{sec:eval_time}). The same seed must not be used for both types of generators~\cite{ksw-gpngi-22}, since typical PRNGs are invertible, hence, their outputs disclose information about the initial value.
Similarly to PRNGs, our general purpose seed generator is invertible. Consequently, this seed can disclose information about the initial PUF response. Hence, cryptographic seed- and key generators should never operate on a memory region that was used by the simple seeder before.

\paragraph{Secure Seeds on Soft Reset}
A fresh and secure seed that was used on CSPRNG initialization should be disguised after use to preserve privacy. Hence, we hash it after CSPRNG seeding and keep the updated value in memory, for a future soft-reset. This prevents backtracking of former random sequences. A future soft-reset adds a soft-reset counter and re-hashes it.
This provides statistical variation among soft-resets (general purpose seeds follow that same procedure).
Disclosure of the updated seed, however, makes future sequences predictable.
Hence, a status indication field (using the \textit{.noinit} PUF attribute) can report the PUF status persistently. CSPRNG initialization can follow its own policy to accept or reject seeding after soft-reset.

\subsection{Key Generation}\label{sec:integ_keygen}

\begin{figure}
    \centering
    \subfloat[Enrollment]{\includegraphics[width=1.0\columnwidth]{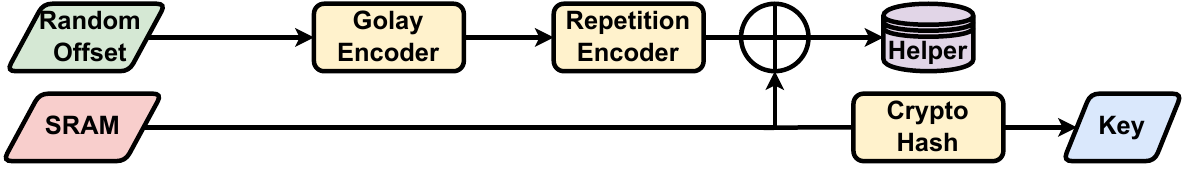}\label{fig:keygen_overview_entroll}}\\
    \subfloat[Reconstruction]{\includegraphics[width=1.0\columnwidth]{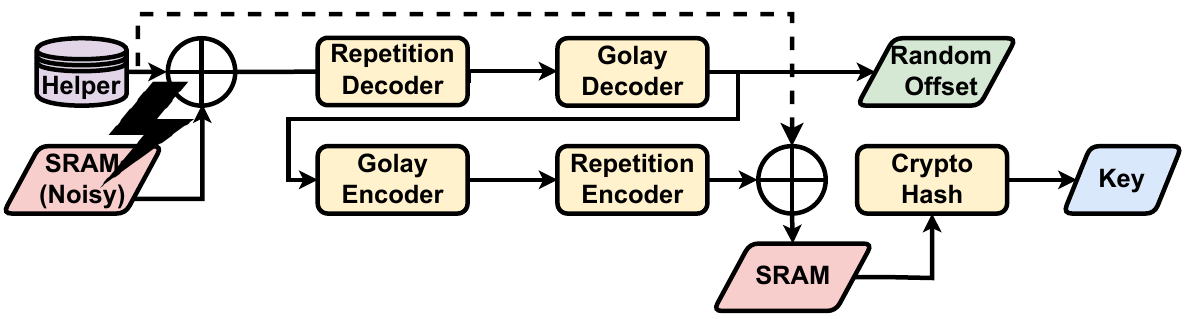}\label{fig:keygen_overview_reconstruct}}
    \caption{A fuzzy extractor based on the code-offset construction. Offset is created at random. Deployments consist of enrollment, and reconstruction during regular device operation.}
    \label{fig:keygen_overview}
\end{figure}

\begin{figure*}[t!]
    \centering
    \begin{adjustbox}{max width=1\textwidth}
        \subfloat[Convergence of the min. entropy estimators for different bias values]{\includegraphics[height=.2\paperheight]{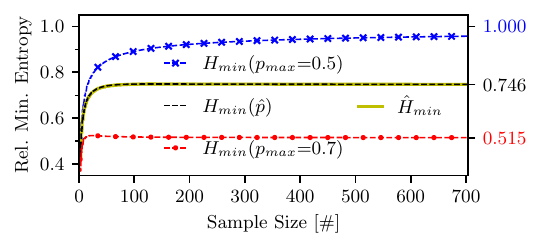}\label{fig:convergence_inter}}
        \subfloat[Standard deviation of min. entropy measurements]{\includegraphics[height=.2\paperheight]{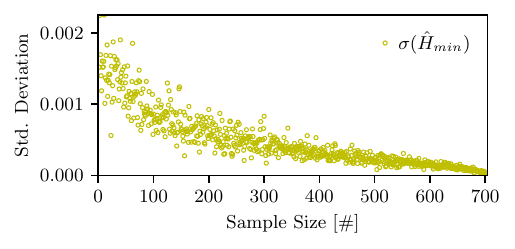}\label{fig:convergence_inter_err}}
    \end{adjustbox}
    \caption{Expectation and measurement of the min. entropy for varying max. probabilities ($p_{max}$) and increasing sample sizes.}
    \label{fig:convergence}
\end{figure*}

Our  key generator follows the approach of the code offset method~\cite{jw-fcs-99}. Deployments of such a system consist of two phases, namely the enrollment (Figure~\autoref{fig:keygen_overview_entroll}), which has to be executed in a trusted environment, and the reconstruction (Figure~\ref{fig:keygen_overview_reconstruct}), which reflects regular device operation.

\paragraph{Enrollment}
Our key generator provides two enrollment options.
\one Helper data is calculated on the device itself.
This greatly simplifies a deployment
and allows for re-enrollment during deployment time (\eg via firmware updates). Re-enrollment must be authenticated, though, to prevent invalidation of intact helper data.
Self-assessment takes a reference measurement utilizing a low-power power-cycle.
A true randomness source is required to generate the random code offset~\cite{jw-fcs-99} (\cf~\autoref{fig:keygen_overview}).
We utilize the PUF based secure seed (see~\autoref{sec:integ_seeder}) to initialize a crypto-secure SHA256PRNG,
which provides unpredictable code offsets of configurable lengths.
\two Helper data is calculated externally, which is convenient for devices with very limited hardware resources. Thereby, a reference SRAM readout is transmitted via UART and an external (trusted) party deals with code offset generation and encoding.
In turn, helper data is formatted into a header file that is part of the subsequent compilation of the firmware.
This option requires individual compilation for every device to deploy.

For the error correction scheme, we rely on lightweight alternatives, namely, a concatenation of the Golay~\cite{g-ndc-49}- and repetition codes, which provide output bit error probabilities of approx. $10^{-5}$ to $10^{-7}$ for common PUF failure rates and lengths~\cite{bgsst-ehdke-08}. Our modular OS integration allows a seamless replacement of corrections codes in the future.

\paragraph{Reconstruction}
A device can reconstruct the key after a power-off cycle, utilizing the helper data. After error correction, the key is calculated by a secure hash (SHA256 by default) and stored in a reserved key variable (see~\autoref{sec:integ_startup}) to prevent overwriting by subsequent OS startup code.
Isolated memory resources are more secure and could hold keys in future, if available on the hardware platform.

\subsection{Access to PUF Primitives}\label{sec:ineg_access}
Random seeds and the secret key are not directly accessible by the user to prevent unauthorized readout, misuse, or tampering.
Instead, this vulnerable data is utilized during module initialization, before application code starts in \texttt{main}.
Seeds are consumed on (CS)PRNG initialization and further processed to obfuscate secret start values, as well as to prepare for the case of a future soft-reset (\autoref{sec:integ_seeder}). As a result, application code simply faces a readily usable (CS)PRNG.
The secret key can be utilized for the initialization of consuming modules, \eg as a master key for deriving additional keys that bootstrap security protocols, or to decrypt secured storage. By the end of the module initialization, the PUF derived key is erased, to prevent direct access by the \texttt{main} application. Hence, it does not persist through a soft-reset but 
requires a real power-off cycle to be re-generated. Alternatively, the key can be stored in isolated memory with controlled access in the future.
 \section{Evaluation of OS-integrated SRAM PUFs}\label{sec:eval_cond}

\subsection{Estimation of the Min. Entropy Convergence}\label{sec:eval_entropy}

\paragraph{Bitwise Inter-device Minimal Entropy}
We want to evaluate the unpredictability of uninitialized SRAM between multiple devices using the min. entropy.
\one Based on experiment data, we measure the relative frequency $p_{max}=max(p, 1-p)$ for attaining one ($p$) or zero ($1-p$)  at the same SRAM bit position of the different devices.
Based on a vector of $p_{max}$ values for every bit position, we evaluate the empirical min. entropy for varying sample sizes (\cf~\autoref{eq:hmin_i_n})
For this, we pick ten sets of devices randomly, and calculate their average min. entropy and $p$-values.
\two An estimator theoretically calculates the expected min. entropy or the empirical estimator as a function of the sample size, \ie the number of nodes, and the maximum probability for logical zero or one.

\paragraph{Robustness of Estimator}
To assess the validity of our min. entropy measurements, we evaluate its convergence rate. We compare our measurements with a sequence of perfect Bernoulli trials and quantify the convergence for different values of $p_{max}$ (\cf~\autoref{sec:back_large}).

Figure~\autoref{fig:convergence_inter} presents the results with convergence limits labeled at the right y-axis. For different $p_{max}$ values, the estimated convergence rate varies.
Exemplary, a $p_{max}$ of 0.7 decreases the number of samples needed for convergence, but it also decreases the relative min. entropy $H_{min}(p_{max}=0.7)$ down to $\approx$\,0.5.
In contrast, the ideal case of $p_{max}=0.5$ should converge to $H_{min}(p_{max}=0.5)$ $\approx$\,1, which however does not occur within 700 displayed samples. This demonstrates the need for large sample sizes.

In our measurements, we find a relative frequency of $\hat{p}_1=0.596$, which slowly converges to a  min. entropy of $\hat{H}_{min} \approx 0.749$ after more than 125 samples.
The standard deviation of our measurements $\sigma(\hat{H}_{min})$ yields $2.3\cdot10^{-3}$ at max. (Figure~\ref{fig:convergence_inter_err}), and decreases with increasing sample sizes.
A comparison of measurement results with our empirical estimator shows almost perfect agreement.
We conclude that our measurements with 708 nodes are empirically robust.

\subsection{Blockwise Evaluation of the Uniqueness}\label{sec:eval_mem_block}
\begin{figure*}
    \centering
    \begin{adjustbox}{max width=1\textwidth}
        \subfloat[Between 708 nodes.]{\includegraphics[height=.2\paperheight]{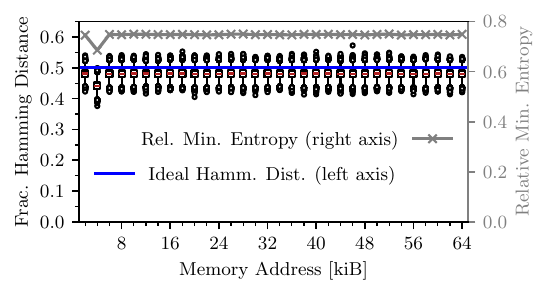}\label{fig:inter_dists_hamm_blocks1024}}
        \subfloat[Between $\approx$ 700 reboots on one device.]{\includegraphics[height=.2\paperheight]{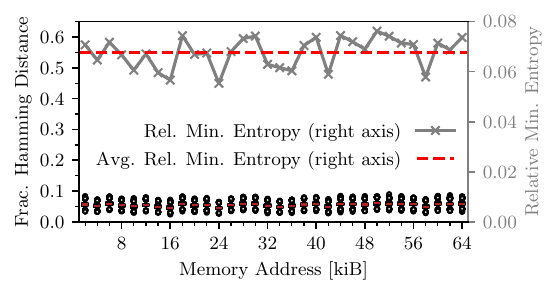}\label{fig:intra_dists_hamm_blocks1024}}
    \end{adjustbox}
    \caption{SRAM evaluation. A total of 64\,kB SRAM is split and analyzed in blocks of 1024\,Bytes. Boxes show fractional hamming distances between all blocks (left y-axis) and lines show min. entropies (right y-axis). IQR: 25th--75th percentile, whiskers: Q1-1.5$\cdot$IQR and Q3+15$\cdot$IQR.}
    \label{fig:dists_hamm_blocks1024}
\end{figure*}

\paragraph{Evaluation between Devices}
We want to quantify the device uniqueness and analyze the fractional hamming distance~\cite{h-edecc-50} between devices and blocks, as a preparation to derive unpredictable secrets:

\begin{equation}
    HD(r_a, r_b) = | \{r_{a, i} \neq r_{b, i} : 1 \le i \le m\} | \cdot \frac{1}{m}
    \label{eq:distance}
\end{equation}

\noindent  where $r_{a,i}$ and $r_{b,i}$ denote the bit values of two devices at position $i$ in a block of the length $m$=1024\,Bytes.
Figure\autoref{fig:inter_dists_hamm_blocks1024} displays our results for the fractional hamming distance between unique device pairs.
Assuming a location-independent occurrence of zeros and ones, the ideal distance is 0.5.
Our measurements fluctuate around an average value of 0.48 except for the block at 4\,kB (bootloader, \cf~\autoref{sec:eval_mem_bias}), a slight deviation from the optimum case.
Based on these results, we consider memory pattern as unique.

Figure\autoref{fig:inter_dists_hamm_blocks1024} additionally presents the blockwise min. entropies (\cf~\autoref{eq:hmin}) between all devices as a lower bound  of its uniqueness. The min. entropy is commonly used to determine input lengths in crypto-contexts (\eg key lengths). Our results reveal a min. entropy of $\approx$ 75\,\% for each block, which is in agreement with~\autoref{sec:eval_entropy} and sufficient to derive unique secrets.
As an example, a naive key generator would require 171\,Bits of uninitialized memory to create a 128\,Bit maximum entropy key.

\paragraph{Evaluation on a Single Device}
We apply the same methodology to $\approx$ 700 readouts on the same device to quantify its initial randomness, required to derive distinct seeds. Thereby we utilize a low-power cycle with a sleep delay of one second.
Figure~\autoref{fig:intra_dists_hamm_blocks1024} presents our results for the blockwise hamming distances and min. entropies.
The intra-device hamming distances reveal a different picture than the inter-device analysis. Even though a majority of bits remain stable over retries, a small portion adds noise, which leads to intra-device distances of $\approx$ 0.06 (average). This behavior remains stable among all memory blocks.
Bit flips lead to an intra-device min. entropy of 6.8\,\%$\pm 0.51 (\sigma)$, which supports seed generation. Conversely, a reproducible key generator must eliminate these. To dimension sufficient correction schemes, we also search for the bit error probability in every block and between all measurements, and find the maximum at $p_e$=0.086.

\section{Analysis of Seed and Key Generation}\label{sec:eval_seed_key}

\subsection{Analysis of Random Seeds}\label{sec:eval_seeder}
We evaluate the quality of seeding and generate two seeds on each startup, \one a secure 256\,Bit seed with maximum entropy, \two a 32\,Bit general purpose seed for non security purposes.
Our evaluation program triggers periodic power-off cycles of 1\,sec. over two days, which results in $\approx$\,180\,k values per device, and 45.1\,Mbit secure / 5.7\,Mbit general purpose seed bits.
\\
\paragraph{Secure Seeds}
We calculate the required bits from SRAM based on the intra-device min. entropy of $\approx$ 7\,\%  as obtained in \autoref{sec:eval_mem_block}.
 We account for the entropy loss using the leftover hash lemma~\cite{bdkp-lhlr-11} ($L=log_2(1/\epsilon$) with $\epsilon=2^{-256}$ close to uniform) while targeting at 256\,Bit entropy in our final seeds. This requires a minimum of 7314\,Bits/914\,Bytes of uninitialized memory. We conservatively chose 1024\,Bytes.
It is worth noting that SRAM portions should be chosen based on a deployment specific initial evaluation of SRAM properties.
All seed values are unique and uniformly distributed due to the properties of the SHA256 hash.

\begin{figure}[htb!]
    \centering
    \includegraphics[width=1.0\columnwidth]{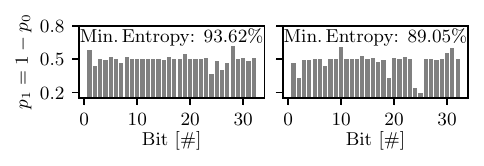}
    \caption{Evaluation of general purpose seeds. Index based distribution of bit probabilities ($p$) throughout 32\,Bit integer values; min. entropy across $\approx$ 180\,k measurements per device.}
    \label{fig:gpseeds}
\end{figure}

\paragraph{General Purpose Seeds}
A min. entropy of 7\,\% requires a minimum of 457\,Bits/57\,Bytes of SRAM to provide 32\,Bit of seed entropy. Conservatively, we choose 128\,Bytes with well aligned values in return.
\autoref{fig:gpseeds} presents the probabilities of $p$ for every bit in the 32\,Bit seed, from two sample devices.
They roughly follow a normal distribution and provide 89--95\,\% min. entropies, which we consider sufficient for non-security purposes.

\subsection{Analysis of the Fuzzy Extractor for Key Generation}\label{sec:eval_keygen}
\begin{figure}[]
    \centering
    \includegraphics[width=.95\columnwidth]{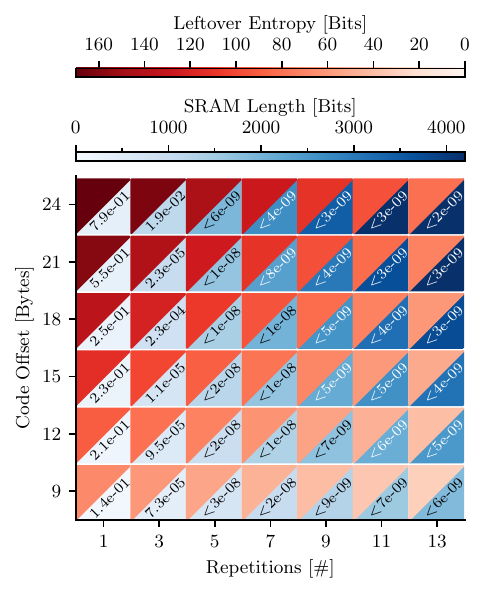}
    \caption{SRAM length, remaining entropy, and measured reconstruction failure rate for different configurations of the fuzzy extractor.}
    \label{fig:len_err_matrix}
\end{figure}

\autoref{fig:len_err_matrix} visualizes the fuzzy extractor properties for varying configurations  (\cf~\autoref{sec:integ_keygen}). Similar to the seed evaluation, every configuration produces $\approx$ 180\,k values.
We vary the code offset from 9 to 24\,Bytes on the y-axis, and the number of repetitions by the repetition error-correction code between 1--13 on the x-axis. A repetition of 1 reflects a single occurrence of the code word. The Golay code is active in all cases.
Lower right triangles in~\autoref{fig:len_err_matrix} (blue) encode the length of required SRAM bits.
The same length is required for helper data on non-volatile memory.

\paragraph{Remaining Key Entropy}
A na{\"i}ve estimation of the entropy of a key output would multiply the SRAM length by the inter-device min. entropy to determine its  cryptographic strength.
For example, a code offset of 9\,Bytes with repetition 1 leads to an SRAM length of 18\,Bytes/144\,Bits;  multiplied with a min. entropy of $\approx$ 0.75 would then yield 108\,Bits of entropy in the SRAM used for key derivation.
For biased SRAM, however, publicly available helper data leak information about the generated key~\cite{klrw-elpkg-14,dgsv-hdapk-15,mlsw-skgbf-16}, which is due to the concatenation of the two error correction codes as part of the fuzzy extractor. This leakage further reduces the remaining entropy in the key and requires additional random code offset- and SRAM bits to compensate.
Maes~\etal~\cite{mlsw-skgbf-16} derived methods for calculating the leakage and the remaining entropy as a function of bias, which reflects the average-case resistance against brute force attacks~\cite{m-ge-94}.

We determine the remaining entropy for varying fuzzy extractor configurations and for our measured SRAM bias of $\hat{p}_1=0.596$ (\cf~\autoref{sec:eval_entropy}).
\autoref{fig:len_err_matrix} visualizes the results for various configurations of the fuzzy extractor. The upper left triangles  (red) reflect the remaining key entropy after fuzzy extraction.
Increasing code offsets increase the required SRAM length (\ie initial entropy) and the remaining entropy in the extracted keys.
Increasing repetitions unsurprisingly increase the required input lengths too, whereas the remaining entropy shows a reversed trend and increases with fewer repetitions.
Code offsets of 24\,Bytes expose remaining entropies from 182\,Bits (1 repetition) down to 82\,Bits (13 repetitions). A random code offset of 24\,Bytes with 5 repetitions provides 144\,Bits of remaining entropy and meets the recommended security strength~\cite{nist-sp80057-2020} of 128\,Bits key entropy.

\paragraph{Reliability}
\autoref{fig:len_err_matrix} also presents the empirical reconstruction failure rate, which is introduced by bit errors between SRAM readouts that cannot be corrected by the fuzzy extractor.
Increasing code offsets increases the error rate (notable in~\autoref{fig:len_err_matrix} following repetitions 1 and 3 for bottom to top).
Following repetitions fewer than five, all fuzzy extractor configurations reveal a notable failure probability, which contradicts the common key reconstruction error rate of $10^{-6}$~\cite{gkst-fiptu-07,lps-sdecc-12,mlsw-skgbf-16,hks-recpe-20}.
Five or more Repetitions expose errors smaller than $3\cdot10^{-8}$. In our measurements, no uncorrected bit error occurred in reconstructed outputs and the error values represent the multiplicative inverse of all successfully reconstructed bits.

\paragraph{Discussion}
Increasing the SRAM length is undesirable since memory is sparse on very constrained IoT devices.
Repetitions should remain few to avoid entropy loss. Conversely, multiple repetitions are required to provide an acceptable  reconstruction rate, in particular for deployments of large SRAM noise level~\cite{sl-casmu-12}.
A code offset of 24\,Bytes and five repetitions preserves sufficient key entropy on our \mthree nodes at a failure rate that meets the requirements for a PUF design.
Other fuzzy extractor configurations either sacrifice reliability by an intolerable reconstruction failure rate at the required level of security, or they sacrifice the remaining key entropy. Our overall balanced strategy provides highly unique and reliable device identities at an acceptable security level.
Pre-processing of the SRAM pattern  as proposed in~\cite{xrfhs-bsash-14,rhgcf-sccna-17,llltl-mecim-18} can further reduce the required SRAM length and increase the remaining key entropy from biased SRAM PUFs, which promises to improve the performance at the same or better security strength.

\subsection{Resource Overhead}\label{sec:eval_time}

\paragraph{Processing Time}
We measure processing times
on \mthree nodes and compare the PUF performance with two different off-the-shelf IoT platforms: \esp and \hifive (see~\autoref{sec:setup_hw}).

First, we analyze the startup latency of two RIOT applications executed on the \mthree node, without the PUF module. \one \texttt{Hello world} is a minimal single-threaded application and introduces a startup latency of 1.1\,ms. \two \texttt{gnrc\_networking} is the standard IPv6 networking application which initializes many modules in 8 threads prior to execution of application code. This requires 10.8\,ms for startup. The latter case excludes seed generation. Here, we utilize a static CSPRNG seed, since microcontroller lacks an entropy source (without the PUF).

\autoref{tab:time_seeder} presents the processing overhead of \one soft-reset detection, \two common routines, and \three both seed generators.
Soft-reset detection is mandatory with our PUF module and adds a
small overhead of < 8\,$\mu$s on the \mthree node. \esp adds $\approx$\,23\,$\mu$s and \hifive surprisingly requires $\approx$\,65 times longer than \mthree.
This is an effect of PUF operation prior to system clock initialization.
Common processing adds a negligible overhead on all platforms.
General purpose seed generation ($\approx$ 0.02--0.13\,ms) is lean
compared to secure seed generation (up to 14\,ms on \mthree) which is comparable to \texttt{gnrc\_networking}, though, seed generation from real entropy sources is slow in general~\cite{ksw-gpngi-22}.
Secure seeds take  $\pm$\,12\,ms on \esp and \hifive. Flash memory access during SHA256 computation is slower on \hifive due to a serial interface.
In agreement with previous measurements, processing times do not directly reflect CPU frequency.
Initializing clocks prior to PUF execution can improve performance in the future, but requires rearrangement of the OS startup routine.
Exemplary, we rearrange the startup code for the \mthree platform and find a speedup of almost 7 times, though, system clock speed increased by a factor of 9, comparing the hardware default state (8\,MHz) and the RIOT configuration (72\,MHz).

\begin{table}[]
    \setlength{\tabcolsep}{4pt}
    \begin{center}
        \caption{Additional operating system startup latencies introduced by soft-reset detection and generation of two seeds.}
        \label{tab:time_seeder}
        \begin{adjustbox}{max width=1\columnwidth}
            \begin{tabularx}{\columnwidth}{rrccr}
            \toprule
            \multirow{3}{1cm}{\textbf{Platform}}
            &
\multirow{3}{1.9cm}{\centering \textbf{Soft-reset detection [ms]}}
            &
\multirow{3}{2cm}{\centering \textbf{Common routines [ms]}}
            &
            \multicolumn{2}{c}{\textbf{Seed generation [ms]}} \\
            \cmidrule{4-5}
            &
            &
            &
            \textbf{simple} &
            \textbf{secure}
            \\
\midrule
\mthree & $7.65\cdot10^{-3}$    & $3.00\cdot10^{-3}$ & 0.13 & 13.65 \\
            \esp    & $22.59\cdot10^{-3}$   & $0.47\cdot10^{-3}$ & 0.02 & 1.37 \\
            \hifive & $494.91\cdot10^{-3}$  & $1.50\cdot10^{-3}$ & 0.08 & 27.03 \\
            \bottomrule
            \end{tabularx}
        \end{adjustbox}
    \end{center}
\end{table}

\begin{figure}
    \centering
    \hfill{\includegraphics[width=1\columnwidth]{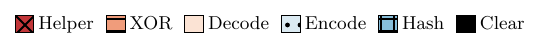}\label{fig:legend_time_fuzzy}}\vspace{-.5cm}\\
\subfloat[Reps.:\,5, Code offs.:\,24\,Bytes]{\includegraphics[width=.49\columnwidth]{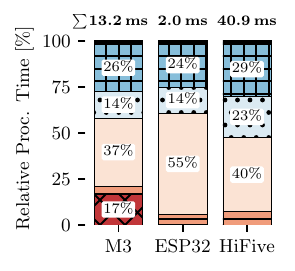}\label{fig:time_fuzzy3}}
    \subfloat[Reps.:\,9, Code offs.:\,24\,Bytes]{\includegraphics[width=.49\columnwidth]{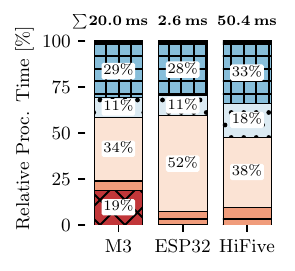}\label{fig:time_fuzzy4}}\\
    \subfloat[Reps.:\,5, Code offs.:\,18\,Bytes]{\includegraphics[width=.49\columnwidth]{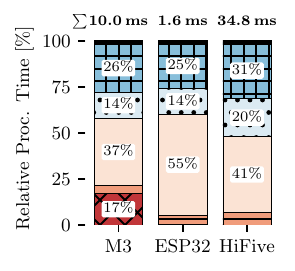}\label{fig:time_fuzzy1}}
    \subfloat[Reps.:\,9, Code offs.:\,18\,Bytes]{\includegraphics[width=.49\columnwidth]{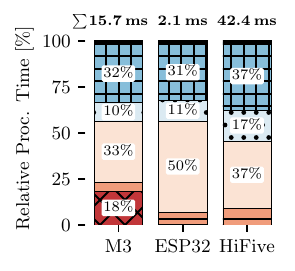}\label{fig:time_fuzzy2}}
    \caption{Additional OS startup latency introduced by PUF reconstruction for four configurations of the fuzzy extractor on different boards. Unlabeled bars relate to proportions below \,10\%.}
    \label{fig:time_fuzzy}
\end{figure}

Next, we look at the processing overhead of the fuzzy extractor and focus on reconstruction since enrollments happen rarely.
We present four relevant configurations for key construction in~\autoref{fig:time_fuzzy}.
`Helper' contains readout of the helper data from flash.
`XOR' contains the overhead from bitwise xor operation at the input and output of the fuzzy extractor (Figure~\autoref{fig:keygen_overview_entroll}).
`Decode' includes overhead of the concatenated Golay- and repetition decoder, and `Encode' includes renewed encoding of the corrected code offset.
`Hash' calculates a digest over the reconstructed PUF measurement.
Finally, `Clear' contains the overhead of re-setting vulnerable data structures after usage.

The absolute latency (numbers above bars) depends on the SRAM length and requires 10--20\,ms on \mthree in all presented cases. The order of magnitude compares to \texttt{gnrc\_networking} and the secure seed generator. Reconstruction and seed generation add to the existing startup latency, though.
Other platforms reflect results from~\autoref{tab:time_seeder} and take 1.6--2.6\,ms (\esp) and 35--50\,ms (\hifive) respectively.
Readout of the helper data is only notable on the \mthree ($\approx$\,17\,\%) due to its slow NOR flash.
The relative processing time for fuzzy extraction increases almost linearly with longer code offsets (\autoref{fig:time_fuzzy} bottom to top).
Increasing the number of repetitions (\autoref{fig:time_fuzzy} left to right) also increases the relative hashing time for a reduction in decoding.
Longer inputs affect the cryptographic hash efforts moderately more than the simple decoder.

In summary, the collection of PUF features moderately delays the startup routines of our sample applications. This motivates our modular design, which allows for selective configuration of PUF features. Furthermore, a positive soft-reset detection skips parts of the PUF execution.
The order of tens of milliseconds is still small compared to the required SRAM power-off time (1 second has proven suitable for different platforms) to generate a fresh memory pattern.
In practice, most IoT applications only awake a few times per hour or day, which obviates the latency overhead.

\paragraph{Energy Consumption}
\autoref{tab:energy_startup} presents the average current flow and the energy consumption of PUF execution on the \mthree node, including the generation of two seeds following four configurations of our fuzzy extractor, in agreement with~\autoref{fig:time_fuzzy}.
To compare these results, we have added a \textit{Comparative case} which acts as a simple alternative for seed- and key inclusion without the PUF. Thereby, we create two seeds by requesting noise from the external radio module (without further conditioning) on the \mthree board and assume a pre-provisioned key to persist in external flash memory, which we import. The quality of randomness, unpredictability, and uniqueness, in this case does not compare to the PUF.
PUF execution prior to systems initialization drains a small current of less than 8.3\,mA on average, whereas the energy consumption ranges from 680--942\,$\mu$J, depending on the input length. This is in line with the processing time.
Our \textit{Comparative case} drains more than five times higher current compared to the PUF, for two reasons. First, this case additionally requires the radio module to be powered. Second, it is operated after systems initialization. This speeds up the execution time ($\approx$3\,ms), however, a higher clock speed further increases the MCU current, leading to a total energy consumption of $\approx$ 440\,$\mu$J.
In summary, the energy demands of the PUF are on the same order of magnitude compared to a simplified use case, and increase the consumption by a factor between 1.5 and 2. Therefore, our PUF contributes conditioned seeds and uniform keys across devices, at little current drain, without depending on additional external hardware modules on the boards.

\begin{table}[]
    \setlength{\tabcolsep}{4pt}
    \begin{center}
        \caption{Current- and energy consumption of the PUF for four configurations of the fuzzy extractor, and a comparative use case---measured on an \mthree node.}
        \label{tab:energy_startup}
        \begin{adjustbox}{max width=1\columnwidth}
            \begin{tabularx}{\columnwidth}{rrrrr|c}
            \toprule
\textbf{Reps. [\#]}    & \textbf{5}&\textbf{9} &\textbf{5} &\textbf{9}&\multirow{2}{*}{\centering\textit{\makecell[c]{Comparative\\case}}}\\
            \textbf{Code offs. [B]}&\textbf{24}&\textbf{24}&\textbf{18}&\textbf{18}&\\
            \midrule
            \textbf{Avg. Current [mA]}     &8.26    &8.19    & 8.28   &8.07    & \textit{44.72}\\
            \textbf{Energy [$\mu$J]  }     &757.49  &941.88  &681.06  &807.21  & \textit{439.52}\\
            \bottomrule
            \end{tabularx}
        \end{adjustbox}
    \end{center}
\end{table}
 \section{Security Analysis}\label{sec:secu_consider}

\newcommand{\tabitem}{~~\llap{\textbullet}~~}
\begin{table*}[]
    \small
    \setlength{\tabcolsep}{4pt}
    \begin{center}
        \caption{Threat overview of the SRAM PUF integration.}
        \begin{tabularx}{\textwidth}{cp{3.4cm}ccccl}
            \toprule
            \textbf{No.}&\makecell[l]{\textbf{Threat description}} & \makecell[c]{\textbf{Asset}\\(\S\ref{sec:secu_assets})} & \makecell[c]{\textbf{Adversary}\\(\S\ref{sec:secu_attacker})}&\makecell[c]{\textbf{Surface}\\(\S\ref{sec:secu_surface})} & \makecell[c]{\textbf{STRIDE}\\(\S\ref{sec:secu_threats})}&  \makecell[c]{\textbf{Mitigation}} \\
            \midrule
T0
            & Readout public helper data.
            & \makecell[c]{A5}
            & \makecell[c]{Hardware}
            & \makecell[t]{S2}
            & \makecell[t]{I}
            & \makecell[tl]{\tabitem Conserv. entropy estim. during enrollment.}\\
            \cmidrule(lr){1-7}
            T1
            & Read/write data.
            & \makecell[c]{A1--5}
            & \makecell[c]{Hardware}
            & \makecell[t]{S2}
            & \makecell[t]{STRIDE}
            & \makecell[tl]{
                \tabitem Enable debug port lock.\\
                \tabitem Use one-time program. memory / write protect.\\
                \tabitem Cut input/output connections.\\
                \tabitem Deploy device with tamper protect. enclosure.}\\
\cmidrule(lr){1-7}
            T2
            & Manipulate operational conditions.
            & \makecell[c]{A1}
            & \makecell[c]{Hardware}
            & \makecell[t]{S3}
            & \makecell[t]{TID}
            & \makecell[tl]{
                    \tabitem Additional entropy sources for seed generation.\\
                    \tabitem Sensors to monitor environ. conditions~\cite{adgk-dfads-21}.
                    } \\
\midrule
T3
            & (Crypto-)analysis of network traffic.
            & \makecell[c]{A1--3}
            & \makecell[c]{Software}
            & \makecell[t]{S1}
            & \makecell[t]{I}
            & \makecell[tl]{
                    \tabitem Conserv. entropy estim. during enrollment.\\
                    \tabitem Short error correction codes.\\
                    } \\
            \cmidrule(lr){1-7}
            T4
            & Readout public/secret data.
            & \makecell[c]{A5}
            & \makecell[c]{Software}
            & \makecell[t]{S1}
            & \makecell[t]{I}
            & \makecell[tl]{
                \tabitem Clear memory after usage.\\
                \tabitem Separate mem. for non-/secure seeds and key.\\
            }\\
            \cmidrule(lr){1-7}
            T5
            & Overwrite control data.
            & \makecell[c]{A4--5}
            & \makecell[c]{Software}
            & \makecell[t]{S1}
            & \makecell[t]{TRID}
            & \makecell[tl]{
                \tabitem Enable hardware assisted soft-reset detection.\\
}\\
            \cmidrule(lr){1-7}
            T6
            & Control operational conditions.
            & \makecell[c]{A1-3}
            & \multirow{2}{*}{\makecell[c]{Software\\(\&Hardware)}}
            & \makecell[t]{S1}
            & \makecell[t]{TID}
            & \makecell[tl]{
                \tabitem Enable hardware assisted voltage detection.\\
}\\
            \bottomrule
        \end{tabularx}
        \label{tab:threat_model}
\end{center}
\end{table*}

PUFs need to maintain unpredictability and unclonability. A secret is embedded in the chip, hence, no seed or key is stored during device sleep, the prevalent state of a battery-driven IoT device.
Secrets only persist during a short time after system startup, reducing the attack vector to a limited time. Practical attacks, however, can still exploit a number of vectors.
We identify \one assets, \two attackers, and \three attack surfaces of our PUF module, and present \four threats.
Risks arise from the combination of specific hardware capabilities, the deployment consideration, application requirements, and related attacker assumptions.
Hence, we are aiming to provide an overview of the prevalent risks, together with a series of mitigations.

\subsection{Assets}\label{sec:secu_assets}
The most vulnerable resources of the SRAM PUF are the uninitialized memory pattern (\textbf{A1}), the output of the PUF, namely secure seeds (\textbf{A2}), and the key (\textbf{A3}). These assets must preserve confidentiality and integrity.
In our implementation, the memory marker (\textbf{A4}) (\eg for soft-reset detection, see~\autoref{sec:integration-riot}) persist after OS startup and is vulnerable because it controls the next reset behavior, \ie can instruct to skip or execute the PUF on a future reset. Hence, this data must preserve integrity.
Non-volatile memory~(\textbf{A5}) stores helper data that is required for key reconstruction. Although helper data is considered public, it is still susceptible.
It must preserve integrity and availability to reconstruct the PUF correctly. Authenticity is also desired, but conventionally very challenging to achieve.

\subsection{Adversaries}\label{sec:secu_attacker}
We distinguish two types of adversaries.
First, \textbf{software} attackers that try to compromise, manipulate, or analyze the system under attack without hardware access. This includes crypto-analysis and the application of learning algorithms.
Software attackers exploit software backdoors, weak implementations, or software bugs to reveal secret information, or disturb code execution. Considering networked nodes in the IoT, attackers can be in wireless reach or connected remotely.
Second, \textbf{hardware} attackers that have direct physical device access. We distinguish two types of hardware attackers:
\textit{Non-invasive} attackers try to interface the device during sleep or operation. They utilize interfaces such as system peripherals, or try to manipulate the device operation conditions.
\textit{Invasive} hardware attackers have deep knowledge and access to advanced techniques to gather or manipulate information on the silicon level. We exclude \textit{invasive} attacks from the reminder of this section because they are \one rare due to high financial and knowledge requirements and \two very specific to chip constructions, and so are mitigations, which contradicts our goal to improve the security of cheap, heterogeneous, and possibly already deployed devices.

\subsection{Surfaces}\label{sec:secu_surface}
We categorize the attack surfaces into three groups.
\one The communication interface (\textbf{S1}), \eg the low-power radio can act as an entry point to inject malicious inputs, or be used for (crypto-) analysis of protocols that make use of random numbers derived by the PUF seed, or the key derived by the fuzzy extractor. This interface also acts as entry point for software updates (future work).
\two I/Os provide an interface to the MCU  (\textbf{S2}). Peripherals such as UART, SPI, or GPIO can revel system internals through logging output, and open an attack vector for interaction with the system.
More crucial, debugging interfaces such as JTAG open a direct interface to the chip memory.
\three The physical presence of a device (\textbf{S3}) provides a surface to operational conditions (\eg temperature, magnetic field) and the power supply.

\subsection{Threats \& Mitigations}\label{sec:secu_threats}
We classify threats using STRIDE~\cite{kg-top-99} which
defines six categories of security threats: Spoofing identity (\textbf{S}), Tampering with data (\textbf{T}), Repudiation (\textbf{R}), Information disclosure (\textbf{I}), Denial of service (\textbf{D}), and Elevation of privilege (\textbf{E}). \autoref{tab:threat_model} summarizes our results and presents mitigations for hardware (\textbf{T0--T2}) and software~(\textbf{T3--T6}) adversaries.

\paragraph{T0} An attacker manages to read non-volatile memory,
by (physically) connecting to the flash memory.
Without the PUF, persistent keys would be stored as plain text, directly disclosing the secret.
PUFs provide additional security by storing only the public helper data in flash. This attack, however, may disclose information in cases of high bias. Hence, helper data readout should still be impractical.

\paragraph{T1} An attacker manages to read/write data such as the uninitialized SRAM pattern, seeds, or keys. Debug interfaces can directly interact with the processor.
Adversaries that manage to connect to the debug lines and initiate a debug session, can halt the CPU during startup to read out memory. If the PUF primitive is used for authentication, this enables spoofing and elevation of privileges without repudiation.
Tampering can invalidate operation leading to denial of service which, however, is simple to achieve with physical device access.
It is noteworthy that PUFs do not introduce additional threats compared to pure software solutions.

\paragraph{T2} An attacker manages to tamper by manipulating environmental operation conditions of the device. Common examples vary the temperature or control the power supply, \eg the power-off time, operation voltage, or startup slope.
This affects random physical processes, including but not limited to SRAM startup state.
A reduction of entropy disqualifies seeds and discloses information, especially in combination with T0.
False key reconstruction can lead to denial of service.
Without the PUF, applications require alternative sources for seed generation, or sometimes use TRNGs permanently which are similarly affected by the environment.
PUFs thus act as an additional source of entropy to increase seed security.

\paragraph{T3} An attacker monitors (secured) network traffic that utilizes random numbers or keys.
This attack might be complemented by owning and analyzing the SRAM on a device of the same type, exploiting bias to predict the initial SRAM pattern.
Crypto-analysis of the output of known algorithms can disclose information of secrets derived from insufficient entropy.
Combined with T0, learning attacks become a risk ~\cite{sfp-mlpuf-21} in these cases.
Without the PUF, however, random numbers are unavailable on platforms without a TRNG which fully prevents security.
Keys are sometimes shipped by the vendor and reutilized across devices, leading to zero entropy on large quantities of nodes~\cite{rswo-igncz-17,thn-midus-15}.
PUFs enable security contributing a uniformly random seed and a unique key that is derived from individual device variations.

\paragraph{T4} An attacker manages to read data structures through software backdoors, which challenges privacy regardless of PUFs (\eg compromise of keys in working memory).
At the time that the network interface is up and running, SRAM is not uninitialized anymore, and vulnerable seeds should be cleared.
A state compromise of a non-forward secure PRNG, however, potentially allows backtracking of the initial SRAM pattern and discloses information.
Without the PUF, initial secrets are likely stored persistently in plain text.
PUFs reduce this attack surface to helper data disclosure (see T0).

\paragraph{T5} An attacker manages to overwrite vulnerable data structures (\eg forcing buffer overflow). Tampering with the soft-reset memory marker (\autoref{sec:integ_softreset}) can trigger a false negative detection on next reset, which leads to zero entropy seeding and defect key reconstruction. Similarly, tampering helper data is a risk. This causes information disclosure and enables denial of service without repudiation. Interfering with code execution threatens code execution regardless of PUFs, though.

\paragraph{T6} Combines T2 and T5. An attacker manages to tamper with the voltage supply through \textbf{software} interfaces--likely present in low-power OSes for undervolting.
Dynamic adjustments during program execution that do not affect startup conditions after reset (sleep) are uncritical, since the PUF is processed before operation.
Adjustments that persist after reset, however, are crucial. A lower voltage causes the reduction of SRAM entropy which disqualifies seeds, leading to information disclosure. Alternative random sources might be similarly affected by this attack (see T2).

\paragraph{Threat discussion}
PUFs provide non-uniform keys across devices, which means that each device has to be attacked separately, rather than attacking one and owning all devices.
The success of a hardware attacker depends on \one the device accessibility and \two the additional security features of the chip.
Hardware attacks are typically small-scale, which contradicts the large-scale characteristics of common IoT deployments.
A \textit{non-invasive} hardware attack requires high efforts for a single device, whereas many threats can be mitigated by standard hardware features.
High security applications, however, should design specific hardware-security features and consider device enclosures.

Software attacks are more likely in the IoT since devices become accessible remotely through the network. Thereby, the attack surface reduces notably, compared to hardware attacks.
Presuming an adequate enrollment, the prevalent software-based threat is given by information disclosure and tampering through a software backdoor (T4--T6).
These threats, however, do not assault the PUF in particular, but generally impede operation of this constrained device class. Hence, the PUF adds a layer of security in practice.
To reduce this attack surface, vendors include trusted execution environments (\eg ARM TrustZone, RISC-V PMP) on modern IoT platforms, that allow for code isolation and privileged memory access. Privileged PUF operations can improve security by separating user facing, networking, or driver code from secure operations.
Conversely, secure processing environments require a root of trust, which can be assisted with a PUF. Hence, both features could complement each other in the future. \section{Conclusion and Outlook}\label{sec:conclusion-outlook}

This paper started from the observation that many commodity IoT devices provide little to no hardware security features, sometimes not even  a source of randomness.
We presented the first comprehensive PUF integration into an  IoT operating system to fill this gap and broadly enhance embedded security.
Our PUF proposal uses uninitialized SRAM, which is available on common IoT platforms, and is portable due to an integration below the hardware abstraction layer of the open-source operating system.

We evaluated SRAM PUF on typical class~2~devices in an open testbed using 708 nodes. This is, to the best of our knowledge, the first empirical PUF study with several hundreds constrained IoT nodes, albeit prior work~\cite{wp-cbmpu-19} proved the need for large sample sizes for the subtle analysis of SRAM bias. 
Our analysis revealed four key insights.

\one An inter-device distance of $\approx$\,48\,\% between node pairs shows high uniqueness, which enables the generation of unpredictable keys.
Still, the physical SRAM layout introduces inter-device bias, which becomes visible when analyzing high numbers of nodes. This reduces the inter-device min. entropy to $\approx$\,75\,\%, and thereby the number of unpredictable bits per node. 
Key generation relies on public helper data, which may reveal information about the SRAM pattern in the case of bias.
Our analysis of the entropy leakage identified a fuzzy extractor configuration that results in 144\,Bits of remaining key entropy at a failure rate of $6\cdot10^{-9}$.
\two An intra-device min. entropy of $\approx$\,7\,\% allows for secure seed generation on startup.
\three The uninitialized SRAM properties of real-word aged, heavily utilized testbed nodes are still sufficient to achieve \one and \two.
\four A configurable OS integration can seamlessly provide PUF services to the IoT at moderate start-up overhead while shielding soft resets.

We could also  show that a number of hardware-based \textit{non-invasive} attacks against SRAM PUFs heavily depend on the availability of platform features such as device pinouts or debug port locks. The availability of PUFs upgrades the security of commercial off-the-shelf devices without cryptographic hardware and strengthens the resistance against the more dangerous software attacks from remote parties throughout the Internet.
Contributing non-uniform keys across devices, our PUF integration reduces the efficacy of these attacks, since each node needs to be attacked
individually, rather than attacking one and owning all.

This work opens four future research directions.
First, pre-processing of biased SRAM pattern may increase the security of keys while reducing the fuzzy extractor overhead, but it adds a layer of complexity to the generation process, which needs careful evaluation on resource constrained devices.
Second, an aging detection and an anti-aging stage may observe and mitigate entropy loss on degrading nodes.
Third, the PUF functions can be extended  to include trusted execution environments, which become increasingly available on modern hardware.
Fourth, integrated analysis tools may improve estimates of entropy and SRAM length. We hope  this will ease deployment efforts toward a future, more secure IoT.
 
\paragraph{Acknowledgments}
We would like to thank Nils Wisiol for his careful feedback, which has significantly helped to improve the paper.
This work was supported in part by the German Federal Ministry for Education and Research (BMBF) within the project \textit{PIVOT: Privacy-Integrated design and Validation in the constrained IoT.}

\paragraph{Availability of software and reproducibility}
We  support reproducible research \cite{acmrep,swgsc-terrc-17} and utilize open source software
and open testbed platforms. All of our work is  publicly released.
The code of the software components, pre-compiled binary images, the implementation of the estimator, documentation, data sets and related tools are available on GitHub at \url{https://github.com/inetrg/IEEE-TDSC-PUF23}.
 \label{lastbodypage}
\bibliographystyle{IEEEtran}
\bibliography{own,rfcs,ids,ngi,iot,layer2,meta,complexity,internet,theory,security,programming}

\begin{thebibliography}{10}
\providecommand{\url}[1]{#1}
\csname url@samestyle\endcsname
\providecommand{\newblock}{\relax}
\providecommand{\bibinfo}[2]{#2}
\providecommand{\BIBentrySTDinterwordspacing}{\spaceskip=0pt\relax}
\providecommand{\BIBentryALTinterwordstretchfactor}{4}
\providecommand{\BIBentryALTinterwordspacing}{\spaceskip=\fontdimen2\font plus
\BIBentryALTinterwordstretchfactor\fontdimen3\font minus
  \fontdimen4\font\relax}
\providecommand{\BIBforeignlanguage}[2]{{%
\expandafter\ifx\csname l@#1\endcsname\relax
\typeout{** WARNING: IEEEtran.bst: No hyphenation pattern has been}%
\typeout{** loaded for the language `#1'. Using the pattern for}%
\typeout{** the default language instead.}%
\else
\language=\csname l@#1\endcsname
\fi
#2}}
\providecommand{\BIBdecl}{\relax}
\BIBdecl

\bibitem{sl-casmu-12}
G.-J. Schrijen and V.~van~der Leest, ``{Comparative analysis of SRAM memories
  used as PUF primitives},'' in \emph{DATE '12: Design, Automation Test in
  Europe Conference Exhibition}.\hskip 1em plus 0.5em minus 0.4em\relax
  Piscataway, NJ, USA: IEEE, 2012, pp. 1319--1324.

\bibitem{kkrsv-pmfbs-12}
S.~Katzenbeisser, {\"U}.~Kocaba{\c{s}}, V.~Ro{\v{z}}i{\'{c}}, A.-R. Sadeghi,
  and I.~V.~C. Wachsmann, ``{PUFs: Myth, Fact or Busted? A Security Evaluation
  of Physically Unclonable Functions (PUFs) Cast in Silicon},'' in
  \emph{Cryptographic Hardware and Embedded Systems (CHES '12)}, E.~Prouff and
  P.~Schaumont, Eds.\hskip 1em plus 0.5em minus 0.4em\relax Berlin, Heidelberg:
  Springer--Verlag, 2012, pp. 283--301.

\bibitem{clb-csfpt-12}
M.~Claes, V.~van~der Leest, and A.~Braeken, ``{Comparison of SRAM and FF PUF in
  65nm Technology},'' in \emph{Information Security Technology for
  Applications}, P.~Laud, Ed.\hskip 1em plus 0.5em minus 0.4em\relax Berlin,
  Heidelberg: Springer--Verlag, 2012, pp. 47--64.

\bibitem{bbmm-tmspq-15}
M.~Barbareschi, E.~Battista, A.~Mazzeo, and N.~Mazzocca, ``{Testing 90 nm
  microcontroller SRAM PUF quality},'' in \emph{10th International Conference
  on Design \& Technology of Integrated Systems in Nanoscale Era
  (DTIS'15)}.\hskip 1em plus 0.5em minus 0.4em\relax Piscataway, NJ, USA: IEEE,
  2015.

\bibitem{chlms-avrtt-13}
M.~Cortez, S.~Hamdioui, V.~van~der Leest, R.~Maes, and G.-J. Schrijen,
  ``{Adapting voltage ramp-up time for temperature noise reduction on
  memory-based PUFs},'' in \emph{International Symposium on Hardware-Oriented
  Security and Trust (HOST'13)}.\hskip 1em plus 0.5em minus 0.4em\relax
  Piscataway, NJ, USA: IEEE, 2013, pp. 35--40.

\bibitem{ljj-pctrs-19}
J.~Lee, D.-W. Jee, and D.~Jeon, ``{Power-up control techniques for reliable
  SRAM PUF},'' \emph{IEICE Electronics Express}, vol.~16, no.~13, 2019.

\bibitem{lssth-eitrn-12}
V.~van~der Leest, E.~van~der Sluis, G.-J. Schrijen, PimTuyls, and H.~Handschuh,
  \emph{{Efficient Implementation of True Random Number Generator Based on SRAM
  PUFs}}.\hskip 1em plus 0.5em minus 0.4em\relax Berlin, Heidelberg: Springer,
  2012, pp. 300--318.

\bibitem{hbf-pssif-09}
D.~E. Holcomb, W.~P. Burleson, and K.~Fu, ``{Power-Up SRAM State as an
  Identifying Fingerprint and Source of True Random Numbers},'' \emph{IEEE
  Transactions on Computers}, vol.~58, no.~9, pp. 1198--1210, 2009.

\bibitem{ksw-gpngi-22}
\BIBentryALTinterwordspacing
P.~Kietzmann, T.~C. Schmidt, and M.~W{\"a}hlisch, ``{A Guideline on
  Pseudorandom Number Generation (PRNG) in the IoT},'' \emph{ACM Comput.
  Surv.}, vol.~54, no.~6, pp. 112:1--112:38, July 2022. [Online]. Available:
  \url{https://dl.acm.org/doi/10.1145/3453159}
\BIBentrySTDinterwordspacing

\bibitem{kscga-atcai-19}
D.~Kumar, K.~Shen, B.~Case, D.~Garg, G.~Alperovich, D.~Kuznetsov, R.~Gupta, and
  Z.~Durumeric, ``{All Things Considered: An Analysis of IoT Devices on Home
  Networks},'' in \emph{28th {USENIX} Security Symposium ({USENIX} Security
  19)}.\hskip 1em plus 0.5em minus 0.4em\relax Santa Clara, CA: {USENIX}
  Association, Aug. 2019, pp. 1169--1185.

\bibitem{aabbb-umb-17}
M.~Antonakakis, T.~April, M.~Bailey, M.~Bernhard, E.~Bursztein, J.~Cochran,
  Z.~Durumeric, J.~A. Halderman, L.~Invernizzi, M.~Kallitsis, D.~Kumar,
  C.~Lever, Z.~Ma, J.~Mason, D.~Menscher, C.~Seaman, N.~Sullivan, K.~Thomas,
  and Y.~Zhou, ``{Understanding the Mirai Botnet},'' in \emph{26th {USENIX}
  Security Symposium ({USENIX} Security 17)}.\hskip 1em plus 0.5em minus
  0.4em\relax Vancouver, BC: {USENIX} Association, Aug. 2017, pp. 1093--1110.

\bibitem{mcms-lscr-10}
A.~Maiti, J.~Casarona, L.~McHale, and P.~Schaumont, ``{A large scale
  characterization of RO-PUF},'' in \emph{International Symposium on
  Hardware-Oriented Security and Trust (HOST'10)}.\hskip 1em plus 0.5em minus
  0.4em\relax Piscataway, NJ, USA: IEEE, 2010, pp. 94--99.

\bibitem{kks-insrs-06}
S.~V. Kumar, C.~H. Kim, and S.~S. Sapatnekar, ``{Impact of NBTI on SRAM read
  stability and design for reliability},'' in \emph{7th International Symposium
  on Quality Electronic Design (ISQED'06)}.\hskip 1em plus 0.5em minus
  0.4em\relax Los Alamitos, CA, USA: IEEE Computer Society, 2006.

\bibitem{wp-cbmpu-19}
F.~Wilde and M.~Pehl, ``{On the Confidence in Bit-Alias Measurement of Physical
  Unclonable Functions},'' in \emph{International New Circuits and Systems
  Conference (NEWCAS'19)}.\hskip 1em plus 0.5em minus 0.4em\relax Piscataway,
  NJ, USA: IEEE, 2019.

\bibitem{bghkl-rosos-18}
\BIBentryALTinterwordspacing
E.~Baccelli, C.~G{\"u}ndogan, O.~Hahm, P.~Kietzmann, M.~Lenders, H.~Petersen,
  K.~Schleiser, T.~C. Schmidt, and M.~W{\"a}hlisch, ``{RIOT: an Open Source
  Operating System for Low-end Embedded Devices in the IoT},'' \emph{IEEE
  Internet of Things Journal}, vol.~5, no.~6, pp. 4428--4440, December 2018.
  [Online]. Available: \url{http://dx.doi.org/10.1109/JIOT.2018.2815038}
\BIBentrySTDinterwordspacing

\bibitem{RFC-7228}
C.~Bormann, M.~Ersue, and A.~Keranen, ``{Terminology for Constrained-Node
  Networks},'' IETF, RFC 7228, May 2014.

\bibitem{thn-crngf-21}
{The Hacker News}, ``{A Critical Random Number Generator Flaw Affects Billions
  of IoT Devices},''
  \url{https://thehackernews.com/2021/08/a-critical-random-number-generator-flaw.html},
  last accessed 29-03-2022, 2021.

\bibitem{prtg-pof-02}
R.~Pappu, B.~Recht, J.~Taylor, and N.~Gershenfeld, ``{Physical One-Way
  Functions},'' \emph{Science}, vol. 297, no. 5589, pp. 2026--2030, 2002.

\bibitem{gcdd-sprf-02}
B.~Gassend, D.~Clarke, van Marten~Dijk, and S.~Devadas, ``{Silicon Physical
  Random Functions},'' in \emph{Proc. of the 9th ACM Conference on Computer and
  Communications Security (CCS '02)}.\hskip 1em plus 0.5em minus 0.4em\relax
  New York, NY, USA: ACM, 2002, pp. 148--160.

\bibitem{snffi-csitd-18}
M.~G. Samaila, M.~Neto, D.~A.~B. Fernandes, M.~M. Freire, and P.~R.~M.
  In\'{a}cio, ``{Challenges of securing Internet of Things devices: A
  survey},'' \emph{Security and Privacy}, vol.~1, no.~2, p. e20, 2018.

\bibitem{smsak-dpufc-17}
S.~Satpathy, S.~K. Mathew, V.~Suresh, M.~A. Anders, H.~Kaul, A.~Agarwal, S.~K.
  Hsu, G.~Chen, R.~K. Krishnamurthy, and V.~K. De, ``{A 4-fJ/b Delay-Hardened
  Physically Unclonable Function Circuit With Selective Bit Destabilization in
  14-nm Trigate CMOS},'' \emph{IEEE Journal of Solid-State Circuits}, vol.~52,
  no.~4, pp. 940--949, 2017.

\bibitem{tra-iutmp-22}
S.~Taneja, V.~K. Rajanna, and M.~Alioto, ``{In-Memory Unified TRNG and
  Multi-Bit PUF for Ubiquitous Hardware Security},'' \emph{IEEE Journal of
  Solid-State Circuits}, vol.~57, no.~1, pp. 153--166, 2022.

\bibitem{hlyy-azber-23}
Y.~He, D.~Li, Z.~Yu, and K.~Yang, ``{ASCH-PUF: A ``Zero'' Bit Error Rate CMOS
  Physically Unclonable Function With Dual-Mode Low-Cost Stabilization},''
  \emph{IEEE Journal of Solid-State Circuits}, pp. 1--11, early access, Jan.
  2023.

\bibitem{gkst-fiptu-07}
J.~Guajardo, S.~S.Kumar, G.-J. Schrijen, and P.~Tuyls, ``{FPGA Intrinsic PUFs
  and Their Use for IP Protection},'' in \emph{Cryptographic Hardware and
  Embedded Systems (CHES'07)}, P.~Paillier and I.~Verbauwhede, Eds.\hskip 1em
  plus 0.5em minus 0.4em\relax Berlin, Heidelberg: Springer--Verlag, 2007, pp.
  63--80.

\bibitem{mbwry-pt-19}
T.~McGrath, I.~E. Bagci, Z.~M. Wang, U.~Roedig, and R.~J. Young, ``{A PUF
  taxonomy},'' \emph{Applied Physics Reviews}, vol.~6, no.~1, p. 011303, 2019.

\bibitem{ekl-lrpms-11}
I.~Eichhorn, P.~Koeberl, and V.~van~der Leest, ``{Logically Reconfigurable
  PUFs: Memory-Based Secure Key Storage},'' in \emph{Proc. of the 6th ACM
  Workshop on Scalable Trusted Computing (STC '11)}.\hskip 1em plus 0.5em minus
  0.4em\relax New York, NY, USA: ACM, 2011, pp. 59--64.

\bibitem{ccm-pscpi-17}
U.~Chatterjee, R.~S. Chakraborty, and D.~Mukhopadhyay, ``{A PUF-Based Secure
  Communication Protocol for IoT},'' \emph{ACM Trans. Embed. Comput. Syst.},
  vol.~16, no.~3, 2017.

\bibitem{brr-rcaii-21}
G.~Bianchi, A.~L. Rosa, and G.~Restuccia, ``{RIOT-AKA: cellular-like
  authentication over IoT devices},'' in \emph{29th IEEE Int. Conf. on Network
  Protocols (ICNP'21)}.\hskip 1em plus 0.5em minus 0.4em\relax Piscataway, NJ,
  USA: IEEE, 2021, pp. 1--6.

\bibitem{ffgrs-pstsc-21}
A.~Falcone, C.~Felicetti, A.~Garro, A.~Rullo, and D.~Sacc\`{a}, ``{PUF-Based
  Smart Tags for Supply Chain Management},'' in \emph{16th International
  Conference on Availability, Reliability and Security (ARES'21)}.\hskip 1em
  plus 0.5em minus 0.4em\relax New York, NY, USA: ACM, 2021.

\bibitem{ssw-splra-11}
S.~Schulz, A.-R. Sadeghi, and C.~Wachsmann, ``{Short Paper: Lightweight Remote
  Attestation Using Physical Functions},'' in \emph{Proc. of the 4th ACM
  Conference on Wireless Network Security (WiSec '11)}.\hskip 1em plus 0.5em
  minus 0.4em\relax New York, NY, USA: ACM, 2011, pp. 109--114.

\bibitem{pvb-tfuid-17}
M.~A. Prada-Delgado, A.~V\'{a}zquez-Reyes, and I.~Baturone, ``{Trustworthy
  firmware update for Internet-of-Thing Devices using physical unclonable
  functions},'' in \emph{Global Internet of Things Summit (GIoTS '17)}.\hskip
  1em plus 0.5em minus 0.4em\relax Piscataway, NJ, USA: IEEE, 2017, pp. 1--5.

\bibitem{h-ata-21}
{Marten van Hulst}, ``{Anchoring TrustZone with SRAM PUF},''
  \url{https://community.arm.com/arm-community-blogs/b/architectures-and-processors-blog/posts/anchoring-trustzone-with-sram-puf},
  last accessed 09-29-2021, 2019.

\bibitem{fn-atopd-14}
R.~Faraji and H.~R. Naji, ``{Adaptive Technique for Overcoming Performance
  Degradation Due to Aging on 6T SRAM Cells},'' \emph{IEEE Transactions on
  Device and Materials Reliability}, vol.~14, no.~4, pp. 1031--1040, 2014.

\bibitem{gwhs-drsea-19}
U.~Guin, W.~Wang, C.~Harper, and A.~D. Singh, ``{Detecting Recycled SoCs by
  Exploiting Aging Induced Biases in Memory Cells},'' in \emph{International
  Symposium on Hardware Oriented Security and Trust (HOST'19)}.\hskip 1em plus
  0.5em minus 0.4em\relax Piscataway, NJ, USA: IEEE, 2019, pp. 72--80.

\bibitem{wfp-ebcmp-19}
F.~Wilde, C.~Frisch, and M.~Pehl, ``{Efficient Bound for Conditional
  Min-Entropy of Physical Unclonable Functions Beyond IID},'' in
  \emph{International Workshop on Information Forensics and Security
  (WIFS'19)}.\hskip 1em plus 0.5em minus 0.4em\relax Piscataway, NJ, USA: IEEE,
  2019.

\bibitem{rhgcf-sccna-17}
M.~T. Rahman, A.~Hosey, Z.~Guo, J.~Carroll, D.~Forte, and M.~Tehranipoor,
  ``{Systematic Correlation and Cell Neighborhood Analysis of SRAM PUF for
  Robust and Unique Key Generation},'' \emph{Journal of Hardware and Systems
  Security}, vol.~1, no.~2, pp. 137--155, 2017.

\bibitem{xrfhs-bsash-14}
K.~Xiao, M.~T. Rahman, D.~Forte, Y.~Huang, M.~Su, and M.~Tehranipoor, ``{Bit
  selection algorithm suitable for high-volume production of SRAM-PUF},'' in
  \emph{International Symposium on Hardware-Oriented Security and Trust
  (HOST'14)}.\hskip 1em plus 0.5em minus 0.4em\relax Piscataway, NJ, USA: IEEE,
  2014, pp. 101--106.

\bibitem{kacp-pcspu-19}
A.~R. Korenda, F.~Afghah, B.~Cambou, and C.~Philabaum, ``{A Proof of Concept
  SRAM-based Physically Unclonable Function (PUF) Key Generation Mechanism for
  IoT Devices},'' in \emph{Workshop on Security Trust and Privacy in Emerging
  Cyber-Physical Systems (SECON'19)}.\hskip 1em plus 0.5em minus 0.4em\relax
  Piscataway, NJ, USA: IEEE, 2019.

\bibitem{klrw-elpkg-14}
P.~Koeberl, J.~Li, A.~Rajan, and W.~Wu, ``{Entropy loss in PUF-based key
  generation schemes: The repetition code pitfall},'' in \emph{IEEE
  International Symposium on Hardware-Oriented Security and Trust (HOST
  '14)}.\hskip 1em plus 0.5em minus 0.4em\relax Piscataway, NJ, USA: IEEE,
  2014, pp. 44--49.

\bibitem{f-iptia2-71}
W.~Feller, \emph{{An Introduction to Probability Theory and Its Applications}},
  2nd~ed.\hskip 1em plus 0.5em minus 0.4em\relax New York: Wiley \& Sons, 1971,
  vol.~2.

\bibitem{w-lscsi-17}
F.~Wilde, ``{Large Scale Characterization of SRAM on Infineon XMC
  Microcontrollers as PUF},'' in \emph{Proc. of the 4th Workshop on
  Cryptography and Security in Computing Systems (CS2'17)}.\hskip 1em plus
  0.5em minus 0.4em\relax New York, NY, USA: ACM, 2017, pp. 13--18.

\bibitem{hwgh-lsrpa-18}
R.~Hesselbarth, F.~Wilde, C.~Gu, and N.~Hanley, ``{Large scale RO PUF analysis
  over slice type, evaluation time and temperature on 28nm Xilinx FPGAs},'' in
  \emph{International Symposium on Hardware-Oriented Security and Trust
  (HOST'18)}.\hskip 1em plus 0.5em minus 0.4em\relax Piscataway, NJ, USA: IEEE,
  2018, pp. 126--133.

\bibitem{gclhm-lcesr-20}
C.~Gu, C.-H. Chang, W.~Liu, N.~Hanley, J.~Miskelly, and M.~O'Neill, ``{A
  large-scale comprehensive evaluation of single-slice ring oscillator and
  {PicoPUF} bit cells on 28-nm Xilinx {FPGAs}},'' \emph{Journal of
  Cryptographic Engineering}, vol.~11, no.~3, pp. 227--238, 2021.

\bibitem{yssmd-pmerp-12}
M.-D. Yu, R.~Sowell, A.~Singh, D.~M'Ra\"{i}hi, and S.~Devadas, ``{Performance
  metrics and empirical results of a PUF cryptographic key generation ASIC},''
  in \emph{International Symposium on Hardware-Oriented Security and Trust
  (HOST'12)}.\hskip 1em plus 0.5em minus 0.4em\relax Piscataway, NJ, USA: IEEE,
  2012, pp. 108--115.

\bibitem{hlskv-spsco-13}
A.~van Herrewege, V.~van~der Leest, A.~Schaller, S.~Katzenbeisser, and
  I.~Verbauwhede, ``{Secure PRNG Seeding on Commercial Off-the-shelf
  Microcontrollers},'' in \emph{3rd International Workshop on Trustworthy
  Embedded Devices (TrustED '13)}.\hskip 1em plus 0.5em minus 0.4em\relax New
  York, NY, USA: ACM, 2013, pp. 55--64.

\bibitem{kmg-sswps-17}
K.~Krentz, C.~Meinel, and H.~Graupner, ``{Secure self-seeding with power-up
  {SRAM} states},'' in \emph{ISCC '17: Symposium on Computers and
  Communications}.\hskip 1em plus 0.5em minus 0.4em\relax Heraklion, Greece:
  IEEE, 2017, pp. 1251--1256.

\bibitem{jw-fcs-99}
A.~Juels and M.~Wattenberg, ``{A Fuzzy Commitment Scheme},'' in \emph{Proc. of
  the 6th ACM Conference on Computer and Communications Security (CCS
  '99)}.\hskip 1em plus 0.5em minus 0.4em\relax New York, NY, USA: ACM, 1999,
  pp. 28--36.

\bibitem{dors-fehgs-08}
Y.~Dodis, R.~Ostrovsky, L.~Reyzin, and A.~Smith, ``{Fuzzy Extractors: How to
  Generate Strong Keys from Biometrics and Other Noisy Data},'' \emph{SIAM
  Journal on Computing}, vol.~38, no.~1, pp. 97--139, 2008.

\bibitem{yd-srecp-10}
M.-D. Yu and S.~Devadas, ``{Secure and robust error correction for physical
  unclonable functions},'' \emph{IEEE Design \& Test of Computers}, vol.~27,
  no.~1, pp. 48--65, 2010.

\bibitem{hks-recpe-20}
M.~Hiller, L.~K{\"u}rzinger, and G.~Sigl, ``{Review of error correction for
  PUFs and evaluation on state-of-the-art FPGAs},'' \emph{Journal of
  Cryptographic Engineering}, vol.~10, no.~3, pp. 229--247, 2020.

\bibitem{lps-sdecc-12}
V.~van~der Leest, B.~Preneel, and E.~van~der Sluis, ``{Soft Decision Error
  Correction for Compact Memory-Based PUFs Using a Single Enrollmen}t,'' in
  \emph{Cryptographic Hardware and Embedded Systems (CHES '12)}, E.~Prouff and
  P.~Schaumont, Eds.\hskip 1em plus 0.5em minus 0.4em\relax Berlin, Heidelberg:
  Springer--Verlag, 2012, pp. 268--282.

\bibitem{dgsv-hdapk-15}
J.~Delvaux, D.~Gu, D.~Schellekens, and I.~Verbauwhede, ``{Helper Data
  Algorithms for PUF-Based Key Generation: Overview and Analysis},'' \emph{IEEE
  Transactions on Computer-Aided Design of Integrated Circuits and Systems},
  vol.~34, no.~6, pp. 889--902, 2015.

\bibitem{ed-pufda-07}
G.~S. Edward and S.~Devadas, ``{Physical Unclonable Functions for Device
  Authentication and Secret Key Generation},'' in \emph{Proc. of the 44th
  Annual Design Automation Conference (DAC '07)}.\hskip 1em plus 0.5em minus
  0.4em\relax New York, NY, USA: ACM, 2007, pp. 9--14.

\bibitem{bgsst-ehdke-08}
C.~B\"{o}sch, J.~Guajardo, A.-R. Sadeghi, J.~Shokrollahi, and P.~Tuyls,
  ``{Efficient Helper Data Key Extractor on FPGAs},'' in \emph{Cryptographic
  Hardware and Embedded Systems - CHES 2008}.\hskip 1em plus 0.5em minus
  0.4em\relax Berlin, Heidelberg: Springer-Verlag, 2008, pp. 181--197.

\bibitem{mhv-pffpc-12}
R.~Maes, A.~V. Herrewege, and I.~Verbauwhede, ``{PUFKY: A Fully Functional
  PUF-Based Cryptographic Key Generator},'' in \emph{Cryptographic Hardware and
  Embedded Systems (CHES '12)}, E.~Prouff and P.~Schaumont, Eds.\hskip 1em plus
  0.5em minus 0.4em\relax Berlin, Heidelberg: Springer--Verlag, 2012, pp.
  302--319.

\bibitem{kblsw-pscli-21}
\BIBentryALTinterwordspacing
P.~Kietzmann, L.~Boeckmann, L.~Lanzieri, T.~C. Schmidt, and M.~W{\"a}hlisch,
  ``{A Performance Study of Crypto-Hardware in the Low-end IoT},'' in
  \emph{International Conference on Embedded Wireless Systems and Networks
  (EWSN'21)}.\hskip 1em plus 0.5em minus 0.4em\relax New York, USA: ACM,
  February 2021. [Online]. Available:
  \url{https://dl.acm.org/doi/10.5555/3451271.3451279}
\BIBentrySTDinterwordspacing

\bibitem{mmdv-srpsc-13}
M.-D.~M. Yu, D.~M'Ra\"{\i}hi, S.~Devadas, and I.~Verbauwhede, ``{Security and
  Reliability Properties of Syndrome Coding Techniques Used in PUF Key
  Generation},'' 2013.

\bibitem{mlsw-skgbf-16}
R.~Maes, V.~van~der Leest, E.~van~der Sluis, and F.~Willems, ``{Secure key
  generation from biased PUFs: extended version},'' \emph{Journal of
  Cryptographic Engineering}, vol.~6, no.~2, pp. 121--137, 2016.

\bibitem{llltl-mecim-18}
H.~Liu, W.~Liu, Z.~Lu, Q.~Tong, and Z.~Liu, ``{Methods for Estimating the
  Convergence of Inter-Chip Min-Entropy of SRAM PUFs},'' \emph{IEEE
  Transactions on Circuits and Systems I: Regular Papers}, vol.~65, no.~2, pp.
  593--605, 2018.

\bibitem{gtfs-smlaa-16}
F.~Ganji, S.~Tajik, F.~F\"{a}{\ss}ler, and J.-P. Seifert, ``{Strong Machine
  Learning Attack Against PUFs with No Mathematical Model},'' in
  \emph{Cryptographic Hardware and Embedded Systems (CHES'16)}.\hskip 1em plus
  0.5em minus 0.4em\relax Berlin, Heidelberg: Springer--Verlag, 2016, pp.
  391--411.

\bibitem{slz-aasp-20}
J.~Shi, Y.~Lu, and J.~Zhang, ``{Approximation Attacks on Strong PUFs},''
  \emph{IEEE Transactions on Computer-Aided Design of Integrated Circuits and
  Systems}, vol.~39, no.~10, pp. 2138--2151, 2020.

\bibitem{wtmsz-nnmaa-22}
N.~Wisiol, B.~Thapaliya, K.~T. Mursi, J.-P. Seifert, and Y.~Zhuang, ``{Neural
  Network Modeling Attacks on Arbiter-PUF-Based Designs},'' \emph{IEEE
  Transactions on Information Forensics and Security}, 2022.

\bibitem{rssdd-mapuf-10}
U.~R\"{u}hrmair, F.~Sehnke, J.~S\"{o}lter, G.~Dror, S.~Devadas, and
  J.~Schmidhuber, ``{Modeling Attacks on Physical Unclonable Functions},'' in
  \emph{Proc. of the 17th ACM Conference on Computer and Communications
  Security (CCS'10)}.\hskip 1em plus 0.5em minus 0.4em\relax New York, NY, USA:
  ACM, 2010, pp. 237--249.

\bibitem{sfp-mlpuf-21}
E.~Strieder, C.~Frisch, and M.~Pehl, ``{Machine Learning of Physical Unclonable
  Functions using Helper Data: Revealing a Pitfall in the Fuzzy Commitment
  Scheme},'' \emph{IACR Transactions on Cryptographic Hardware and Embedded
  Systems (TCHES '21)}, vol. 2021, no.~2, pp. 1--36, 2021.

\bibitem{hbnf-cpuf-13}
C.~Helfmeier, C.~Boit, D.~Nedospasov, and J.-P. Seifert, ``{Cloning Physically
  Unclonable Functions},'' in \emph{IEEE International Symposium on
  Hardware-Oriented Security and Trust (HOST '13)}.\hskip 1em plus 0.5em minus
  0.4em\relax Piscataway, NJ, USA: IEEE, June 2013, pp. 1--6.

\bibitem{zowks-rdspc-16}
S.~Zeitouni, Y.~Oren, C.~Wachsmann, P.~Koeberl, and A.-R. Sadeghi, ``{Remanence
  Decay Side-Channel: The PUF Case},'' \emph{IEEE Transactions on Information
  Forensics and Security}, vol.~11, no.~6, pp. 1106--1116, 2016.

\bibitem{eclipse-iedsr-19}
{Eclipse Foundation}, ``{IoT \& Edge Developer Survey Report},''
  \url{https://outreach.eclipse.foundation/iot-adoption-2019}, last accessed
  03-12-2022, 2019.

\bibitem{kgsw-psgri-18}
\BIBentryALTinterwordspacing
P.~Kietzmann, C.~G{\"u}ndogan, T.~C. Schmidt, and M.~W{\"a}hlisch, ``{A PUF
  Seed Generator for RIOT: Introducing Crypto-Fundamentals to the Wild},'' in
  \emph{Proc. of 16th ACM International Conference on Mobile Systems,
  Applications (MobiSys), Poster Session}.\hskip 1em plus 0.5em minus
  0.4em\relax New York, NY, USA: ACM, June 2018. [Online]. Available:
  \url{https://doi.org/10.1145/3210240.3210805}
\BIBentrySTDinterwordspacing

\bibitem{abfhm-filso-15}
C.~Adjih, E.~Baccelli, E.~Fleury, G.~Harter, N.~Mitton, T.~Noel,
  R.~Pissard-Gibollet, F.~Saint-Marcel, G.~Schreiner, J.~Vandaele, and
  T.~Watteyne, ``{FIT IoT-LAB: A large scale open experimental IoT testbed},''
  in \emph{2015 IEEE 2nd World Forum on Internet of Things (WF-IoT)}.\hskip 1em
  plus 0.5em minus 0.4em\relax Piscataway, NJ, USA: IEEE Press, Dec 2015, pp.
  459--464.

\bibitem{gklp-ncmcm-18}
\BIBentryALTinterwordspacing
C.~G{\"u}ndogan, P.~Kietzmann, M.~Lenders, H.~Petersen, T.~C. Schmidt, and
  M.~W{\"a}hlisch, ``{NDN, CoAP, and MQTT: A Comparative Measurement Study in
  the IoT},'' in \emph{Proc. of 5th ACM Conference on Information-Centric
  Networking (ICN)}.\hskip 1em plus 0.5em minus 0.4em\relax New York, NY, USA:
  ACM, September 2018, pp. 159--171. [Online]. Available:
  \url{https://doi.org/10.1145/3267955.3267967}
\BIBentrySTDinterwordspacing

\bibitem{bklsw-usioi-22}
\BIBentryALTinterwordspacing
L.~Boeckmann, P.~Kietzmann, L.~Lanzieri, T.~C. Schmidt, and M.~W{\"a}hlisch,
  ``{Usable Security for an IoT OS: Integrating the Zoo of Embedded Crypto
  Components Below a Common API},'' in \emph{International Conference on
  Embedded Wireless Systems and Networks (EWSN'22)}.\hskip 1em plus 0.5em minus
  0.4em\relax New York, USA: ACM, October 2022, pp. 84--95. [Online].
  Available: \url{https://dl.acm.org/doi/10.5555/3578948.3578956}
\BIBentrySTDinterwordspacing

\bibitem{linux-kconfig-20}
{The Linux Kernel Development Community}, ``{Kconfig Language},''
  \url{https://www.kernel.org/doc/html/latest/kbuild/kconfig-language.html},
  last accessed 28-09-2020, 2020.

\bibitem{k-acp-09}
D.~E. Knuth, \emph{{The Art of Computer Programming (Second Edition)}}.\hskip
  1em plus 0.5em minus 0.4em\relax Reading, MA, USA: Addison Wesley, 2009.

\bibitem{g-ndc-49}
M.~J.~E. Golay, ``{Notes on Digital Coding},'' \emph{Proc. of the Institute of
  Radio Engineers (IRE '49)}, vol.~37, pp. 657--657, 1949.

\bibitem{h-edecc-50}
R.~W. Hamming, ``{Error detecting and error correcting codes},'' \emph{The Bell
  System Technical Journal}, vol.~29, no.~2, pp. 147--160, 1950.

\bibitem{bdkp-lhlr-11}
B.~Barak, Y.~Dodis, H.~Krawczyk, O.~Pereira, K.~Pietrzak, F.-X. Standaert, and
  Y.~Yu, ``{Leftover Hash Lemma, Revisited},'' in \emph{Advances in Cryptology
  (CRYPTO '11)}, P.~Rogaway, Ed.\hskip 1em plus 0.5em minus 0.4em\relax Berlin,
  Heidelberg: Springer--Verlag, 2011, pp. 1--20.

\bibitem{m-ge-94}
J.~L. Massey, ``{Guessing and entropy},'' in \emph{Proceedings of IEEE
  International Symposium on Information Theory (ISIT'94)}, 1994, p. 204.

\bibitem{nist-sp80057-2020}
E.~Barker, ``{Recommendation for Key Management},'' {National Institute of
  Standards and Technology}, Gaithersburg, MD, US, Tech. Rep. NIST SP 800-57
  Part 1, May 2020.

\bibitem{adgk-dfads-21}
M.~T.~H. Anik, J.-L. Danger, S.~Guilley, and N.~Karimi, ``{Detecting Failures
  and Attacks via Digital Sensors},'' \emph{IEEE Transactions on Computer-Aided
  Design of Integrated Circuits and Systems}, vol.~40, no.~7, pp. 1315--1326,
  2021.

\bibitem{kg-top-99}
\BIBentryALTinterwordspacing
L.~Kohnfelder and P.~Garg, ``{The threats to our products},'' Microsoft, Tech.
  Rep., 1999. [Online]. Available:
  \url{https://adam.shostack.org/microsoft/The-Threats-To-Our-Products.docx}
\BIBentrySTDinterwordspacing

\bibitem{rswo-igncz-17}
E.~Ronen, A.~Shamir, A.-O. Weingarten, and C.~O'Flynn, ``{IoT Goes Nuclear:
  Creating a ZigBee Chain Reaction},'' in \emph{IEEE Symposium on Security and
  Privacy (SP)}.\hskip 1em plus 0.5em minus 0.4em\relax Piscataway, NJ, USA:
  IEEE Press, 2017, pp. 195--212.

\bibitem{thn-midus-15}
{The Hacker News}, ``{Millions of IoT Devices Using Same Hard-Coded CRYPTO
  Keys},'' \url{https://thehackernews.com/2015/11/iot-device-crypto-keys.html},
  last accessed 02-12-2022, 2015.

\bibitem{acmrep}
{ACM}, ``{Result and Artifact Review and Badging},''
  \url{http://acm.org/publications/policies/artifact-review-badging}, Jan.,
  2017.

\bibitem{swgsc-terrc-17}
Q.~Scheitle, M.~W{\"a}hlisch, O.~Gasser, T.~C. Schmidt, and G.~Carle,
  ``{Towards an Ecosystem for Reproducible Research in Computer Networking},''
  in \emph{Proc. of ACM SIGCOMM Reproducibility Workshop}.\hskip 1em plus 0.5em
  minus 0.4em\relax New York, NY, USA: ACM, August 2017, pp. 5--8.

\end{thebibliography}
\label{lastpage}

\IEEEdisplaynontitleabstractindextext

\IEEEpeerreviewmaketitle

\section*{Author Biography}
\nobalance \begin{IEEEbiography}[{\includegraphics[width=1in,height=1.25in,clip,keepaspectratio]{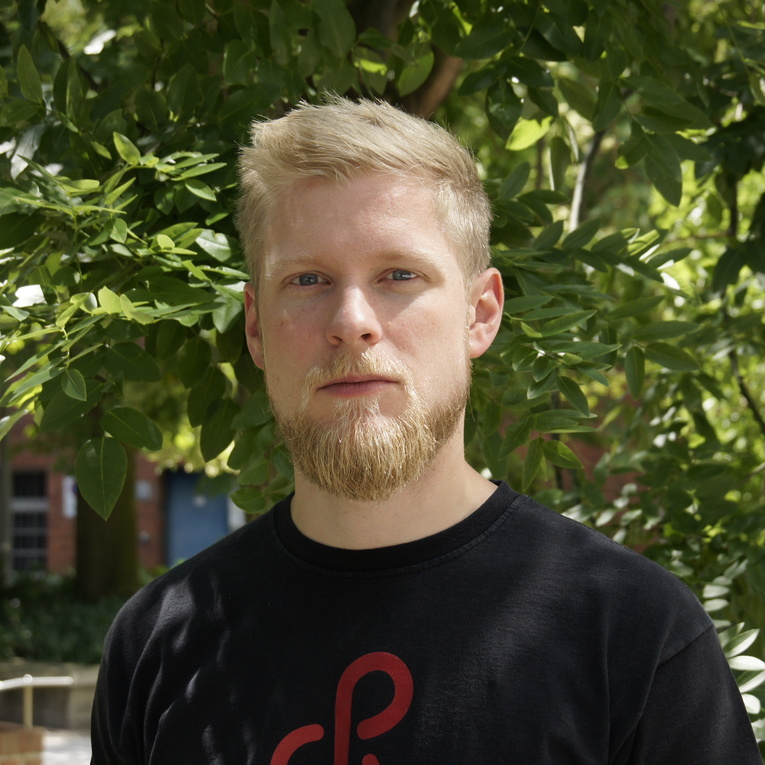}}]{Peter Kietzmann}
is a PhD student of the Internet Technologies research
group at the Hamburg University of Applied Sciences. His particular
interests lie in radio technologies, embedded programming, and secure
IoT protocols. In the German-French research project PIVOT
(Privacy-Integrated design and Validation in the constrained IoT) he is
currently involved, exploring the secure protection of data on low-end
devices and low-power radio networks of the ultra-constrained IoT.
\end{IEEEbiography}

\begin{IEEEbiography}[{\includegraphics[width=1in,height=1.25in,clip,keepaspectratio]{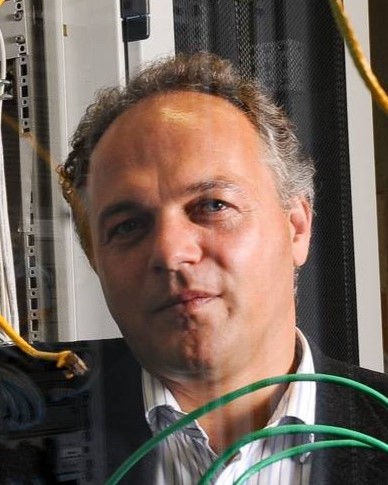}}]{Thomas C. Schmidt}
studied mathematics, physics, and German literature at Freie Universit{\"a}t Berlin (FU Berlin),
Berlin, Germany, and the University of Maryland at College Park, MD. He
received the Ph.D. degree in mathematical physics from FU Berlin, in 1993. He is a Professor of
computer networks and Internet technologies with the Hamburg University of Applied Sciences,
Hamburg, Germany, where he heads the Internet Technologies Research Group. He was the Director of
the HTW Computer Centre, Berlin. Since then, he has continuously conducted numerous national and
international research projects. He was the Principal Investigator in a number of EU, nationally
funded, and industrial projects, as well as a Visiting Professor with the University of
Reading, Reading, U.K. He is also a co-founder and a coordinator of the open source community
developing the RIOT operating system. His current research interests include development,
measurement, and analysis of large-scale distributed systems like the Internet or its offsprings.
Dr. Schmidt has served as a Co-Editor and a Technical Expert on several occasions and is actively
involved in the work of IETF and IRTF, where he
co-chaired the SAM RG.
\end{IEEEbiography}

\begin{IEEEbiography}[{\includegraphics[width=1in,height=1.25in,clip,keepaspectratio]{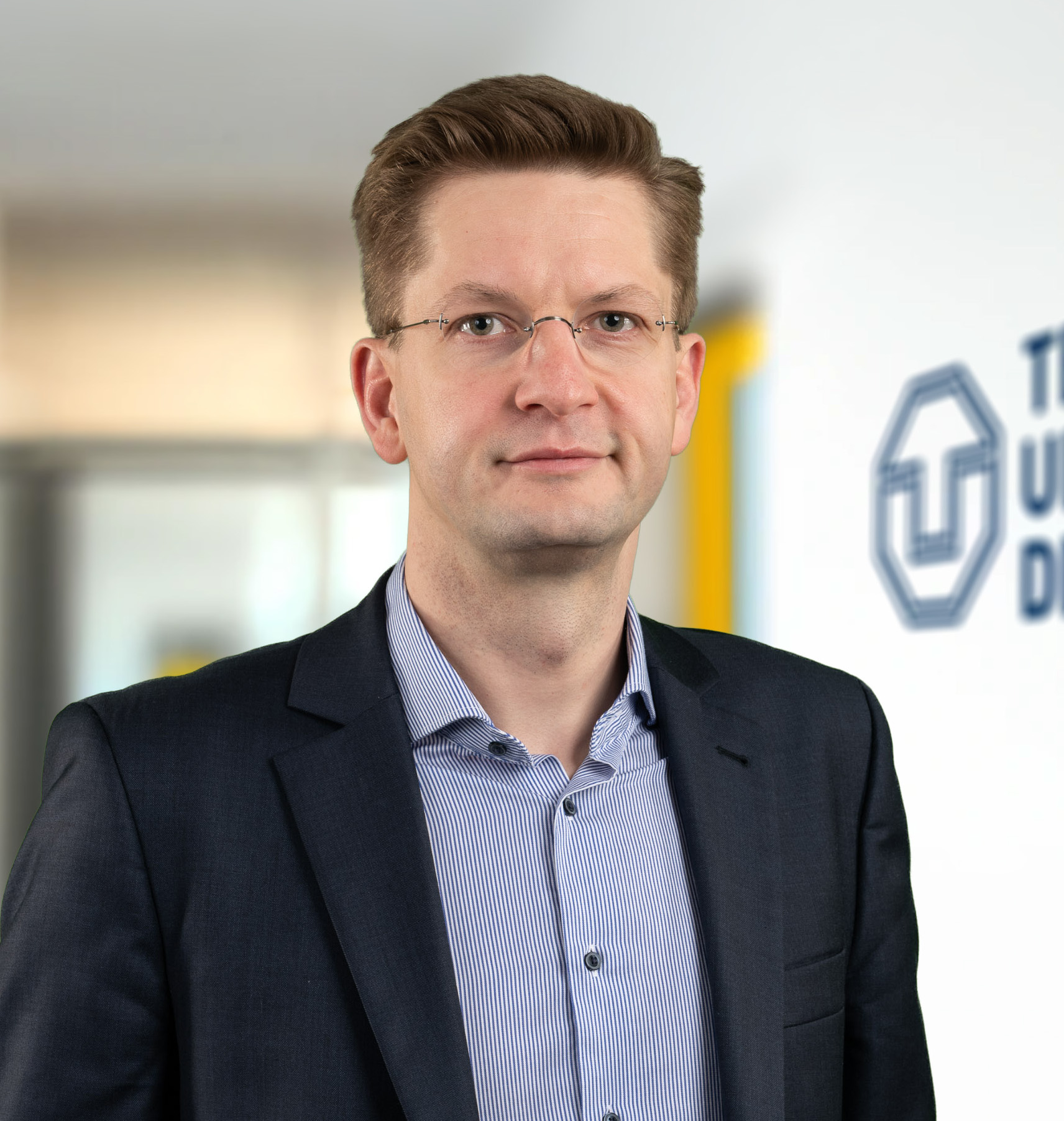}}]{Matthias W{\"a}hlisch}
received the Ph.D. degree (with highest honors) in computer science from
Freie Universit{\"a}t Berlin, Berlin, Germany. He is a full professor and holds the Chair of Distributed and Networked Systems at the Faculty of Computer Science at TU Dresden. He is a co-founder and a coordinator of several successful open
source projects such as RIOT. His efforts are driven by trying to improve Internet communication based on sound research. He is the PI of several national and international projects. He has
authored or co-authored over 130 peer-reviewed papers (e.g., in ACM HotNets, ACM IMC, and ACM
CoNEXT). His current research interests include the design and evaluation of networking protocols and
architectures, as well as Internet measurements and analysis. Dr. W{\"a}hlisch was a recipient of the
Young Talents Award of Leibniz-Kolleg Potsdam for outstanding achievements in advancing the
Internet, and recipient of the Excellent Young Scientists Award for his contributions to the
Internet of Things and their prospective entrepreneurial practice. He has been active in the IETF/
IRTF since 2005. He co-organized or co-chaired over 50 scientific events, including the IEEE ICNP
Ph.D. Forum 2013, ACM IMC 2017, ACM SIGCOMM 2017, and ACM ICN 2022.
\end{IEEEbiography}

\end{document}